\documentstyle[a4p,12pt,epsfig]{article}

\begin{document}
\newcommand {\epem}     {e$^+$e$^-$}
\newcommand {\ecm} { E_{c.m.} }
\newcommand {\mnch} {\langle n_{{\mathrm ch.}} \rangle }
\newcommand {\ptlu} { p_{\perp} }
\newcommand {\ycut} { y_{{\mathrm cut}} }
\newcommand {\ymin} { y_{{\mathrm min}} }
\newcommand {\dymax} { \Delta y_{{\mathrm max}} }
\newcommand {\dygap} { \Delta y_{{\mathrm gap}} }
\newcommand {\qtot} {Q_{{\mathrm leading}}}
\newcommand {\elead} {E_{{\mathrm leading}}}
\newcommand {\mlead} {M_{{\mathrm leading}}}
\newcommand {\twomlead} {M_{{\mathrm leading}}^{+-}}
\newcommand {\mleadzero} {M_{{\mathrm leading}}^{00}}
\newcommand {\fourmlead} {M_{{\mathrm leading}}^{+-+-}}
\newcommand {\ylead} {y_{{\mathrm leading}}}
\newcommand {\nchlead} {n^{{\mathrm ch.}}_{{\mathrm leading}}}
\newcommand {\qjet}  {\kappa_{{\mathrm jet}}}
\newcommand {\delqtot} 
   {\Delta Q_{{\mathrm leading}}^{{\mathrm MC-data}}}
\newcommand {\ptmin} {p_{T,{\mathrm min.}}}
\newcommand {\lamqcd} {\Lambda_{\mathrm QCD}}
\newcommand {\mw} {M_{\mathrm W}}
\def\gsim{\raisebox{-3pt}{\rlap{$\,\sim\,$}} \raisebox{3pt}{$\,>\,$}}

\begin{titlepage}
\noindent
\begin{center}  {\large EUROPEAN ORGANIZATION FOR NUCLEAR RESEARCH }
\end{center}

\begin{tabbing}
\` CERN-EP-2003-031 \\
\` 26 May 2003 \\
\end{tabbing}

\bigskip\bigskip\bigskip\medskip

\begin{center}{\LARGE\bf
Tests of models of color reconnection and \\[2mm]
a search for glueballs using gluon jets
with a rapidity gap
}
\end{center}

\begin{center}
{\Large
The OPAL Collaboration
}
\end{center}

\begin{center}{\large\bf Abstract}\end{center}
\bigskip
\noindent
Gluon jets with a mean energy of 22~GeV
and purity of 95\% are selected
from hadronic Z$^0$ decay events produced in {\epem} annihilations.
A subsample of these jets is identified which
exhibits a large gap in the rapidity distribution of
particles within the jet.
After imposing the requirement of a rapidity gap,
the gluon jet purity is~86\%.
These jets are observed to demonstrate a high degree of
sensitivity to the presence of color reconnection,
i.e.~higher order QCD processes affecting the 
underlying color structure.
We use our data to test three QCD models which include a
simulation of color reconnection:
one in the Ariadne Monte Carlo,
one in the Herwig Monte Carlo,
and the other by Rathsman in the Pythia Monte Carlo.
We find the Rathsman and Ariadne color reconnection
models can describe our gluon jet
measurements only if very large values are used 
for the cutoff parameters which serve 
to terminate the parton showers,
and that the description of inclusive Z$^0$ data
is significantly degraded in this case.
We conclude that color reconnection as implemented
by these two models is disfavored. 
The signal from the Herwig color reconnection model
is less clear and we do not obtain a definite
conclusion concerning this model.
In a separate study,
we follow recent theoretical suggestions
and search for glueball-like objects
in the leading part of the gluon jets.
No clear evidence is observed for these objects.

\vspace*{2cm}
\begin{center}{\large
  (Submitted to Eur. Phys. J. C)
}\end{center}

\end{titlepage}

\begin{center}{\Large        The OPAL Collaboration
}\end{center}\bigskip
\begin{center}{
G.\thinspace Abbiendi$^{  2}$,
C.\thinspace Ainsley$^{  5}$,
P.F.\thinspace {\AA}kesson$^{  3}$,
G.\thinspace Alexander$^{ 22}$,
J.\thinspace Allison$^{ 16}$,
P.\thinspace Amaral$^{  9}$, 
G.\thinspace Anagnostou$^{  1}$,
K.J.\thinspace Anderson$^{  9}$,
S.\thinspace Arcelli$^{  2}$,
S.\thinspace Asai$^{ 23}$
% ...
}\end{center}\bigskip
\bigskip
$^{  1}$School of Physics and Astronomy, University of Birmingham,
Birmingham B15 2TT, UK
\newline
$^{  2}$Dipartimento di Fisica dell' Universit\`a di Bologna and INFN,
I-40126 Bologna, Italy
\newline
$^{  3}$Physikalisches Institut, Universit\"at Bonn,
D-53115 Bonn, Germany
\newline
% ...
$^{  5}$Cavendish Laboratory, Cambridge CB3 0HE, UK
\newline
% ...
$^{  9}$Enrico Fermi Institute and Department of Physics,
University of Chicago, Chicago IL 60637, USA
\newline
% ...
$^{ 16}$Department of Physics, Schuster Laboratory, The University,
Manchester M13 9PL, UK
\newline
% ...
$^{ 22}$Department of Physics and Astronomy, Tel Aviv University,
Tel Aviv 69978, Israel
\newline
$^{ 23}$International Centre for Elementary Particle Physics and
Department of Physics, University of Tokyo, Tokyo 113-0033, and
Kobe University, Kobe 657-8501, Japan
\newline
% ...

\clearpage\newpage

\section{Introduction}
\label{sec-introduction}

Rapidity $y$,
defined by 
$y$$\,=\,$$\frac{1}{2}\ln\left(\frac{E+p_{\parallel}}{E-p_{\parallel}}\right)$
with $E$ the energy of a particle and
$p_{\parallel}$ the component of its 3-momentum 
along an axis\footnote{Usually the
thrust~\cite{bib-thrust}, jet, or beam axis.},
is one of the most common variables used to characterize the
phase space distribution of particles in high energy collisions.
Of current interest (see for example~\cite{bib-bj})
are events with a so-called rapidity gap,
namely events in which two populated regions in rapidity
are separated by an interval devoid of particles.
High energy collisions are often characterized by the formation of 
quark and gluon jets, 
i.e.~collimated streams of hadrons associated with
the hard scattering of quarks and gluons, respectively.
Most recent interest in rapidity gaps has focused on a class of events in
electron-proton~\cite{bib-ep} and proton-antiproton~\cite{bib-pp}
collisions with large rapidity gaps between jets:
these events are interpreted as arising from the
exchange of a strongly interacting color singlet object,
such as a pomeron~\cite{bib-pomeron},
between the underlying partonic constituents
of the event.

\begin{figure} 
 \begin{center}
 \begin{tabular}{cc}
      \epsfxsize=5cm
      \epsffile{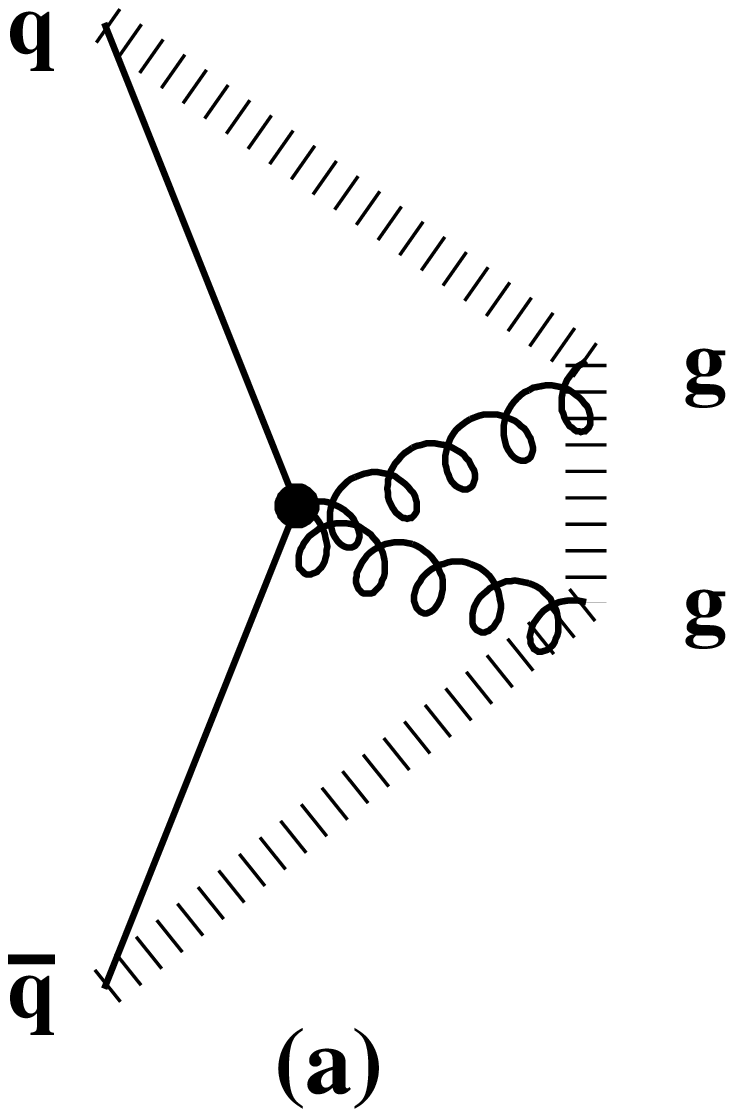} & \hspace*{1.5cm}
      \epsfxsize=5cm
      \epsffile{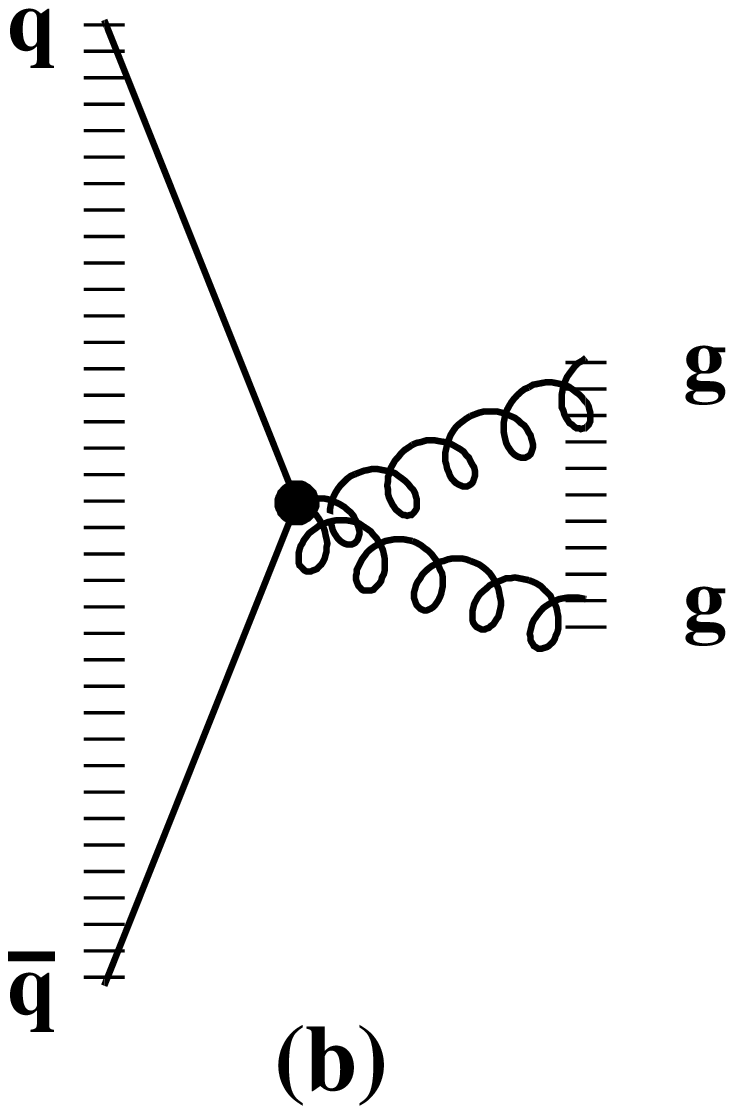}
 \end{tabular}
 \end{center}
\caption{
Schematic illustrations of 
events with (a)~standard ``planar'' color flow and 
(b)~reconnection.
The hatched regions represent color flux tubes or
``strings'' stretched between the quark~q,
antiquark $\mathrm\overline{q}$ and gluons~g.}
\label{fig-planar}
\end{figure}

Another source of rapidity gaps is color reconnection (CR),
i.e.~a rearrangement of the underlying color structure 
of an event from its simplest configuration,
in which a color flux tube or ``string'' is stretched 
from a quark to an antiquark through intermediate gluons
in a manner such that string segments do not cross
(a so-called planar diagram, see Fig.~\ref{fig-planar}a),
to a more complex pattern
in which some segments can either cross
or else appear as disconnected entities whose endpoints
are gluons (Fig.~\ref{fig-planar}b).
Diagrams with color reconnection represent higher order
processes in Quantum Chromodynamics (QCD),
suppressed by order $1/N_{\mathrm C}^2$
compared to planar diagrams,
where $N_{\mathrm C}$$\,=\,$$3$ is the number of colors.
In models of hadron production such as the 
Lund string model~\cite{bib-lund},
the flux tubes hadronize.
In events with a disconnected gluonic string segment
as in Fig.~\ref{fig-planar}b,
a rapidity gap can form between the isolated segment
--~often the leading (highest rapidity) part of a gluon jet~--
and the rest of the event.
Thus rapidity gaps in gluon jets can provide a sensitive means
to search for effects of color reconnection.
Color reconnection has been a topic of considerable
recent interest because 
of its potential effects in
fully hadronic decays of W$^+$W$^-$ events
produced in electron-positron ({\epem}) collisions~\cite{bib-zerwas},
introducing an uncertainty in
the measurement of the W boson mass
at LEP~\cite{bib-opalcr}.

Recently~\cite{bib-ochs},
gluon jets with a rapidity gap were also proposed as
a potentially favorable environment
for the production of color singlet bound states of gluons,
such as glueballs,
through diagrams like Fig.\ref{fig-planar}b in which the isolated
gluonic system represents a hadronic resonance.

Previous studies of
rapidity gaps in {\epem} hadronic annihilations
were based on inclusive Z$^0$ events
and separated two- and three-jet events
from Z$^0$ decays~\cite{bib-sld96}.
The rapidity distribution of charged particles in gluon
jets was used to test models of color reconnection
in~\cite{bib-opalgincl98}.
There are no previously published experimental
studies on gluon jets with a rapidity gap.

In this paper,
we study gluon jets with rapidity gaps,
produced in three-jet quark-antiquark-gluon
($\mathrm q\overline{q}g$) events from 
{\epem} hadronic Z$^0$ decays.
The gluon jets are identified through ``anti-tagging,''
using displaced secondary vertices from
B hadrons to identify the quark and anti-quark jets.
The data were collected with the OPAL detector at the
LEP {\epem} storage ring at CERN.
We measure the charged particle multiplicity,
total electric charge,
and distributions of invariant mass
in the leading part of the gluon jets.

\section{Detector and data sample}
\label{sec-detector}

The OPAL detector is described in
detail elsewhere~\mbox{\cite{bib-detector,bib-si}}.
OPAL operated from 1989 to 2000
and subsequently was dismantled.
The tracking system
consisted of a silicon microvertex detector,
an inner vertex chamber,
a large volume jet chamber,
and specialized chambers at the outer radius of the 
jet chamber to improve the measurements in the
$z$-direction.\footnote{Our right handed
coordinate system is defined so that
$z$~is parallel to the e$^-$ beam axis,
$x$ points towards the center of the LEP ring,
$r$~is the coordinate normal to the beam axis,
$\phi$~is the azimuthal angle around the beam axis with
respect to $x$,
and $\theta$ is the polar angle \mbox{with respect to~$z$.}}
The tracking system covered the region
$|\cos\theta|$$\,<\,$0.98 and
was enclosed by a solenoidal magnet coil
with an axial field of~0.435~T.
Electromagnetic energy was measured by a
lead-glass calorimeter located outside the magnet coil,
which also covered $|\cos\theta|$$\,<\,$0.98.

The present analysis is based on a sample of
about $2\,722\,000$ hadronic annihilation events,
corresponding to the OPAL sample collected
within 3~GeV of the Z$^0$ peak from 1993 to 1995.
This sample includes readout 
of both the $r$-$\phi$ and $z$ 
coordinates of the silicon strip microvertex detector~\cite{bib-si}.
The procedures for identifying hadronic annihilation
events are described in~\cite{bib-opaltkmh}.

We employ the tracks of charged particles
reconstructed in the tracking chambers
and clusters of energy
deposited in the electromagnetic calorimeter.
Charged tracks are required to have at least 20 measured
points (of 159 possible) in the jet chamber,
or at least 50\% of the number of points expected 
based on the track's polar angle,
whichever is larger.
In addition, the tracks are required 
to have a momentum component perpendicular 
to the beam axis greater than 0.05~GeV/$c$,
to lie in the region $|\cos\theta|$$\,<\,$0.96,
to point to the origin to within 5~cm in the $r$-$\phi$ plane
and 30~cm in the $z$ direction,
and to yield a reasonable $\chi^2$ per
degree-of-freedom for the track fit in the $r$-$\phi$ plane.
Electromagnetic clusters are required to have an energy greater 
than 0.10~GeV if they are in the barrel section 
of the detector ($|\cos\theta|$$\,<\,$0.82)
or 0.25~GeV if they are in the endcap section 
(0.82$\,<\,$$|\cos\theta|$$\,<\,$0.98).
A matching algorithm~\cite{bib-mt} is employed to
reduce double counting of energy in cases where 
charged tracks point towards electromagnetic clusters.
Specifically,
if a charged track points towards a cluster,
the cluster's energy is re-defined by subtracting 
the energy which is expected to be deposited in
the calorimeter by the track.
If the energy of the cluster is smaller than 
this expected energy,
the cluster is not used.
In this way,
the energies of the clusters are primarily associated with neutral
particles.

Each accepted track and cluster
is considered to be a particle.
Tracks are assigned the pion mass.
Clusters are assigned zero mass since they originate
mostly from photons.

To eliminate residual background and events
in which a significant number of particles is lost
near the beam direction,
the number of accepted charged tracks in an event
is required to be at least five and the 
thrust axis of the event,
calculated using the particles,
is required to satisfy
$|\cos (\theta_{\mathrm{thrust}})|$$\,<\,$0.90,
where $\theta_{\mathrm{thrust}}$ is the
angle between the thrust and beam axes.
The number of events which passes these cuts is~$2\,407\,000$.
The residual background to this sample
from all sources is estimated to be less 
than~1\%~\cite{bib-opaltkmh} and is neglected.

\section{QCD models}
\label{sec-models}

To establish the sensitivity of our analysis 
to processes with color reconnection,
we generate events using
Monte Carlo simulations of perturbative QCD 
and the hadronization process,
both with and without the effects of reconnection.

The models without reconnection in our study are
the Jetset~\cite{bib-jetset},
Herwig~\cite{bib-herwig,bib-herwig2} and Ariadne~\cite{bib-ariadne}
Monte Carlo programs,
versions 7.4, 6.2 and 4.11 respectively.
Jetset and Herwig are based on parton showers with
branchings described by Altarelli-Parisi splitting functions~\cite{bib-dglap},
followed by string hadronization~\cite{bib-lund} for Jetset
and cluster hadronization~\cite{bib-cluster} for Herwig.
Ariadne employs the dipole cascade model~\cite{bib-dipole}
to generate a parton shower,
followed by string hadronization.
The principal parameters of the models
were tuned to yield an optimized description
of the global properties of hadronic Z$^0$ events
and are documented in~\cite{bib-qg95b} for Jetset
and in Tables~\ref{tab-herwig} and~\ref{tab-ariadne}
for Herwig and Ariadne.
All three models provide a good description of
the main features of {\epem} hadronic annihilation events,
including the properties of identified gluon jets,
see for example~\cite{bib-opalgincl98}.

\begin{table}[p]
\begin{center}
\begin{tabular}{|c|c|c|c|}
\hline
  & & & \\[-2mm] 
  Parameter & Monte Carlo & Default & Optimized \\[-1mm]
            &     name    & value   &   value   \\[2mm]
\hline
  & & & \\[-2mm]
  $\lamqcd$ (GeV)                  & QCDLAM & 0.18 & $0.18\pm0.01$ \\[1mm]
  Gluon mass (GeV/$c^2$)           & RMASS(13) & 0.75 & $0.75\pm 0.05$ \\[1mm]
  Maximum cluster mass parameter (GeV/$c^2$) 
                                   & CLMAX  & 3.35 & $3.35\pm0.05$ \\[1mm]
  Maximum cluster mass parameter   & CLPOW  & 2.00 & $2.0\pm0.2$ \\[1mm]
  Cluster spectrum parameter, udsc & PSPLT(1)  & 1.00 & $1.00\pm0.05$ \\[1mm]
  Cluster spectrum parameter, b 
      & PSPLT(2)  & 1.00 & $0.33^{+0.07}_{-0.03}$ \\[1mm]
  Gaussian smearing parameter, udsc  
      & CLSMR(1)  & 0.0 & $0.40^{+0.20}_{-0.02}$ \\[1mm]
  Decuplet baryon weight & DECWT  & 1.0 & $0.7\pm0.1$ \\[2mm]
\hline
\end{tabular}
\caption{
OPAL parameter set for Herwig, version 6.2.
The method used to tune the parameters 
is presented in~\cite{bib-qg95b}.
Parameters not listed were left at their default values.
The uncertainties represent
$\pm 1$ standard deviation limits obtained from the $\chi^2$
contours.
The $\chi^2$ contours were defined by varying the parameters 
one at a time from their tuned values.
}
\label{tab-herwig}
\end{center}
\vspace*{1cm}
\begin{center}
\begin{tabular}{|c|c|c|c|}
\hline
  & & & \\[-2mm] 
  Parameter & Monte Carlo & Default & Optimized \\[-1mm]
            &     name    & value   &   value   \\[2mm]
\hline
  & & & \\[-2mm]
  $\lamqcd$ (GeV)    & PARA(1)  & 0.22 & $0.215\pm 0.002$ \\[1mm]
  $\ptmin$ (GeV/$c$) & PARA(3)  & 0.60 & $0.70\pm 0.05$ \\[1mm]
  $b$ (GeV$^{-2}$)   & PARJ(42) & 0.58 & $0.63 \pm 0.01$ \\[1mm]
  ${\cal{P}}(qq)/{\cal{P}}(q)$
             & PARJ(1)  & 0.10 & $0.130 \pm 0.003$ \\[1mm]
  $\displaystyle\left[{\cal{P}}(us)/{\cal{P}}(ud) \right]/
    \left[{\cal{P}}(s)/{\cal{P}}(d) \right] $
             & PARJ(3)  & 0.40 & $0.600 \pm 0.016$ \\[1mm]
  $ {\cal{P}}(ud_1)/3{\cal{P}}(ud_0) $
             & PARJ(4)  & 0.05 & $0.040^{+0.010}_{-0.003}$ \\[2mm]
\hline
  & & & \\[-2mm]
  Extra Baryon suppression & PARJ(19) & 1.00 & 0.52 \\
    (MSTJ(12)$\,=\,$3) & & & \\[2mm]
\hline
\end{tabular}
\caption{
OPAL parameter set for Ariadne, version 4.11.
The method used to tune the parameters 
is presented in~\cite{bib-qg95b}.
Parameters not listed were left at their default values.
The extra baryon suppression factor PARJ(19),
enabled by setting MSTJ(12)$\,=\,$3,
was taken from~\cite{bib-alephpr}.
The uncertainties have the same meaning as
in Table~\ref{tab-herwig}.
}
\label{tab-ariadne}
\end{center}
\end{table} 

The models in our study which incorporate color reconnection are
the model of L\"{o}nnblad~\cite{bib-lonnblad}
implemented in the Ariadne Monte Carlo\footnote{There 
are three variants of the color
reconnection model in Ariadne, 
corresponding to settings of the parameter MSTA(35)=1, 2 or 3;
for hard processes involving a single color singlet system,
such as Z$^0$ decays,
all three variants are identical;
note that the parameter PARA(28)
should be set to zero if the MSTA(35)=2 option
is used in Z$^0$ decays~\cite{bib-leif}.},
the color reconnection model~\cite{bib-herwig2} in
the Herwig Monte Carlo,
and a model
introduced by Rathsman~\cite{bib-rathsman}.
We refer to these as the
Ariadne-CR, 
Herwig-CR, and Rathsman-CR models,
respectively.
The Ariadne-CR model is an extension of
the model of Gustafson and H\"{a}kkinen~\cite{bib-gostacr}.
The Rathsman-CR model is implemented in
the Pythia Monte Carlo~\cite{bib-jetset}, version 5.7.
For {\epem} annihilations in the absence of
initial-state photon radiation,
Pythia is equivalent to Jetset.
Thus,
the Rathsman-CR model is effectively a version of
Jetset which contains color reconnection.
We note the Pythia Monte Carlo contains its own color
reconnection model,
based on the work of Khoze and Sj\"{o}strand~\cite{bib-ks}.
We do not include this model in our study because it
is not implemented for Z$^0$ decays.
The Rathsman-CR model has been found to provide a good description of 
rapidity gap measurements in both
electron-proton and proton-antiproton collisions~\cite{bib-rathgood}.

The parameters we use for the Ariadne-CR model are
the same as those given in
Table~\ref{tab-ariadne} for Ariadne
except for the parameter PARJ(42)
which was adjusted from 0.63 to 0.55~GeV$^{-2}$ so that the
model describes the measured value of mean charged particle 
multiplicity in inclusive Z$^0$ decays,
$\mnch$,
see Sect.~\ref{sec-inclusive}.
Analogously,
the parameters of the Herwig-CR model are the
same as those used for Herwig 
(see Table~\ref{tab-herwig})
except CLMAX was adjusted from 3.35 to 3.75~GeV/$c^2$
and RMASS(13) from 0.75 to~0.793~GeV/$c^2$
to describe $\mnch$.
For our implementation of the Rathsman-CR model,
we use the parameter set given for Jetset in~\cite{bib-qg95b}.

Besides the Jetset parameters,
the Rathsman-CR model employs a parameter,
denoted $R_0$,
which is an overall suppression factor
for color reconnection.
The value of $R_0$ is not arbitrary but reflects the
$1/N_{\mathrm C}^2$ suppression of reconnected events
compared to planar events mentioned in the Introduction.
For this parameter,
we use $R_0$$\,=\,$0.1 as suggested in~\cite{bib-rathsman}.
The analogous parameter in the Herwig-CR model,
PRECO,
is maintained at its default value of
$1/N_{\mathrm C}^2$$\,=\,$1/9.
For the Ariadne-CR model,
the corresponding parameter,
PARA(26),
stipulates
the number of distinct dipole color states.
We use the default value for this parameter,
PARA(26)$\,=\,$9,
which again corresponds to $N_{\mathrm C}$$\,=\,$3
and the $1/N_{\mathrm C}^2$ suppression
of reconnected processes.

Our implementations of the Ariadne-CR,
Herwig-CR and Rathsman-CR models
provide descriptions of the global features of {\epem} data
which are essentially equivalent
to those of the corresponding models without reconnection.
This is discussed in Sect.~\ref{sec-inclusive} below.

The Monte Carlo events are examined at two levels:
the ``detector level'' and the ``hadron level.''
The detector level includes initial-state photon radiation,
simulation of the OPAL detector~\cite{bib-gopal},
and the analysis procedures
described in Sect.~\ref{sec-detector}.
The hadron level does not include these effects and utilizes
all charged and neutral particles with lifetimes
greater than 3$\,\times\,$$10^{-10}$~s,
which are treated as stable.
Samples of 6 million Ariadne, Herwig and Jetset events,
and 3 million Ariadne-CR, 
Herwig-CR and Rathsman-CR events,
were processed through the detector simulation and
used as the detector level samples in this study.
The hadron level samples are based on
10~million Monte Carlo events for each model.

\section{Model predictions for inclusive Z$^0$ decays}
\label{sec-inclusive}

The Ariadne-CR, Rathsman-CR and Herwig-CR models yield
descriptions of standard measures of 
properties in inclusive Z$^0$ data
which are essentially equivalent
to those provided by Ariadne, Jetset, and Herwig, respectively,
as stated above.
Thus, color reconnection as implemented 
in these models has only a small effect on the 
global features of inclusive {\epem} events.
To illustrate these points,
we measured the following distributions using the
inclusive Z$^0$ sample discussed in
Sect.~\ref{sec-detector}:
\begin{enumerate}
  \item Sphericity, $S$~\cite{bib-opal1990};
  \item Aplanarity, $A$~\cite{bib-opal1990};
  \item the negative logarithm of the jet resolution scale
    for which an event changes from being classified as a
    three-jet event to a four-jet event, using 
    the Durham jet finder~\cite{bib-durham}, $-\ln (y_{34})$;
  \item charged particle rapidity with respect to the thrust axis,
    $y_{T}$.
\end{enumerate}
Note there are correlations between these
variables and between different bins of some of
the distributions.
Using a sample of Jetset events at the hadron level,
the correlation coefficient between $S$ and $A$
was found to be 0.66,
between $S$ and $-\ln (y_{34})$ $-0.61$,
and between $A$ and $-\ln (y_{34})$ $-0.66$.
Similar results were found using the other models.
The $y_T$ distribution contains one entry per particle,
in contrast to the other distributions which contain one entry
per event.
Therefore, the $y_T$ distribution was not included
in this correlation study.
Taken together,
the four distributions are sensitive to the momentum
structure of an event both in and out of the
three-jet event plane,
to four-jet event structure,
and to particle multiplicity.
They therefore provide a relatively complete and relatively
uncorrelated set of distributions with which to assess
the global features of {\epem} events.

The distributions were corrected to the hadron level
using bin-by-bin factors.
The method of bin-by-bin corrections is described
in~\cite{bib-opal1990}.
Ariadne was used to determine the correction factors.
Ariadne was chosen because it was found to provide a better
description of the data at the detector level than
Jetset or Herwig.
The typical size of the corrections is 10\%.
As systematic uncertainties, we considered the following.
\begin{itemize}
  \item  The other models --~Jetset, Herwig,
    Ariadne-CR, Herwig-CR and Rathsman-CR~-- were used to 
    determine the correction factors, rather than Ariadne. 
  \item  Charged tracks alone were used for the data and
    Monte Carlo samples with detector simulation,
    rather than charged tracks plus electromagnetic clusters.
  \item  The particle selection was further varied, 
    first by restricting charged tracks and electromagnetic clusters
    to the central region of the detector,
    $|\cos\theta|<0.70$,
    rather than $|\cos\theta|<0.96$ for the charged tracks and
    $|\cos\theta|<0.98$ for the clusters, 
    and second by increasing the minimum transverse momentum 
    of charged tracks with respect to the beam axis
    from 0.05~GeV/$c$ to 0.15~GeV/$c$.
\end{itemize}
The differences between the standard results
and those found using each of these conditions
were used to define symmetric systematic uncertainties.
For the first item,
the largest of the described differences
with respect to the standard result
was assigned as the systematic uncertainty,
and similarly for the third item.
The systematic uncertainties were added in quadrature to define
the total systematic uncertainties.
The systematic uncertainty evaluated for each bin was
averaged with the results from its two neighbors to reduce the
effect of bin-to-bin fluctuations.
The single neighbor was used for bins at the ends of
the distributions.

The largest contribution to the
systematic uncertainty of the $S$,
$-\ln (y_{34})$ and $y_T$ distributions 
arose from using the 
Herwig-CR model to correct the data.
For $A$,
the largest systematic effect was from
using Jetset to correct the data.

\begin{figure}[tp]
\begin{center}
  \epsfxsize=15cm
  \epsffile{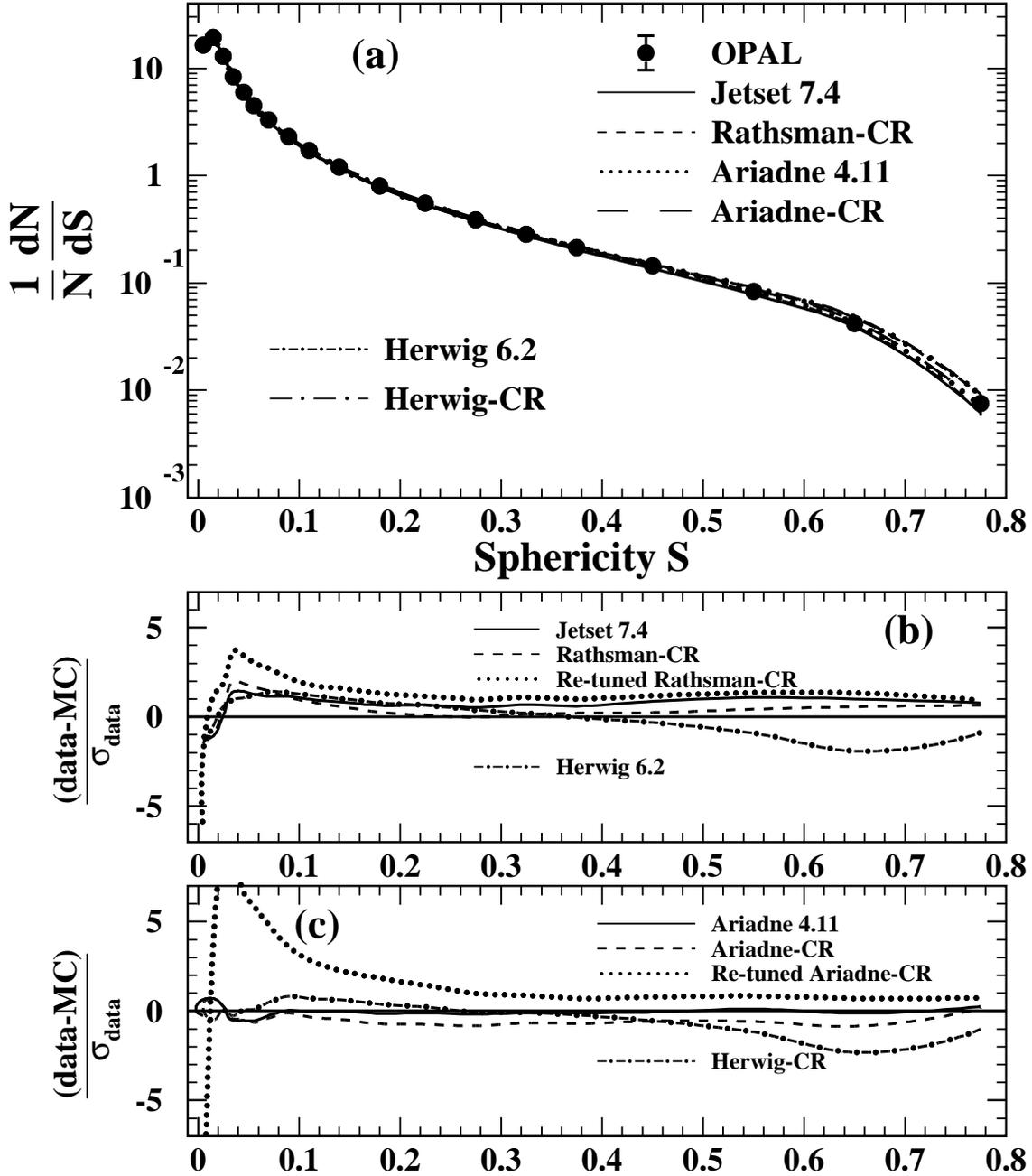}
\end{center}
\caption{
(a) The Sphericity distribution for inclusive Z$^0$ events,
in comparison to
the predictions of models with and without color reconnection (CR).
The data have been corrected for initial-state photon radiation
and detector response.
The statistical uncertainties are too small to be visible.
The vertical lines attached to the data points
(barely visible) show the total uncertainties,
with statistical and systematic terms added in quadrature.
(b)~and~(c) show the deviations of the Monte Carlo predictions
from the data in units of
the total experimental uncertainties, $\sigma_{\mathrm data}$.
}
\label{fig-sphericity}
\end{figure}

\begin{figure}[tp]
\begin{center}
  \epsfxsize=15cm
  \epsffile{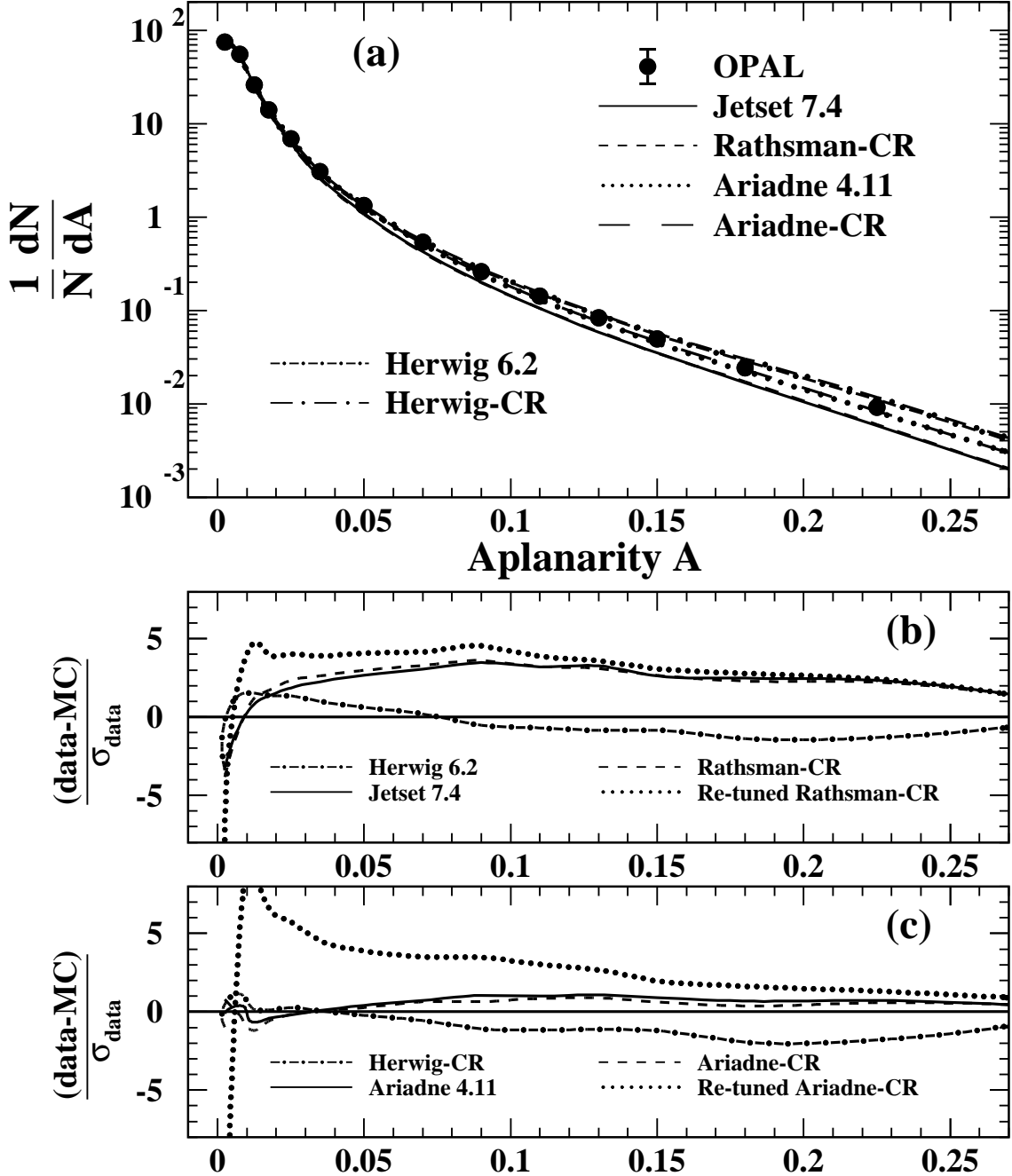}
\end{center}
\caption{
(a) The Aplanarity distribution for inclusive Z$^0$ events,
in comparison to
the predictions of models with and without color reconnection (CR).
The data have been corrected for initial-state photon radiation
and detector response.
The uncertainties --~both statistical and total~--
are too small to be visible.
(b)~and~(c) show the deviations of the Monte Carlo predictions
from the data in units of
the total experimental uncertainties, $\sigma_{\mathrm data}$.
}
\label{fig-aplanarity}
\end{figure}

\begin{figure}[tp]
\begin{center}
  \epsfxsize=15cm
  \epsffile{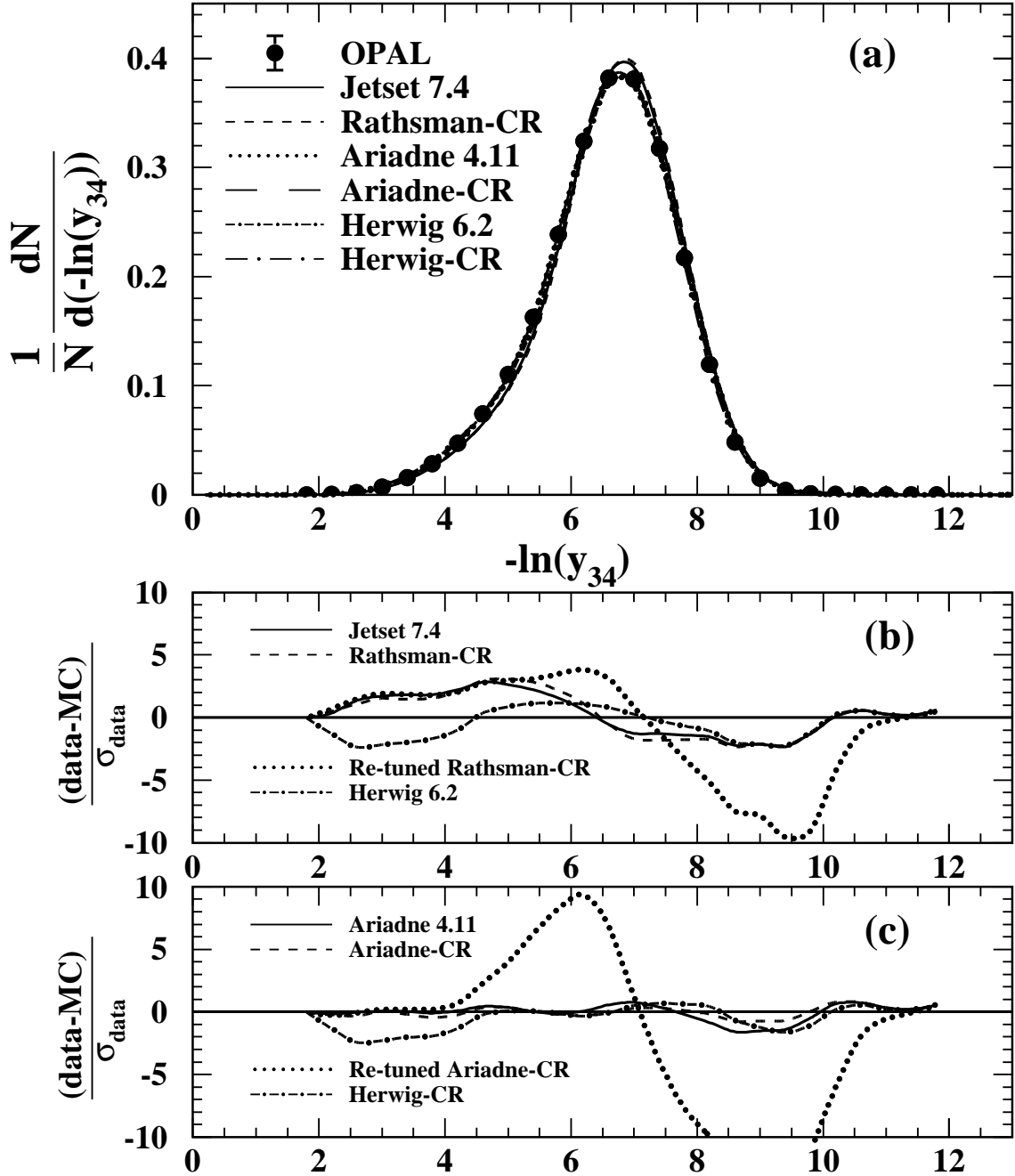}
\end{center}
\caption{
(a) The $-\ln (y_{34})$ distribution for inclusive Z$^0$ events,
in comparison to
the predictions of models with and without color reconnection (CR).
The data have been corrected for initial-state photon radiation
and detector response.
The uncertainties --~both statistical and total~--
are too small to be visible.
(b)~and~(c) show the deviations of the Monte Carlo predictions
from the data in units of
the total experimental uncertainties, $\sigma_{\mathrm data}$.
}
\label{fig-y34}
\end{figure}

\begin{figure}[tp]
\begin{center}
  \epsfxsize=15cm
  \epsffile{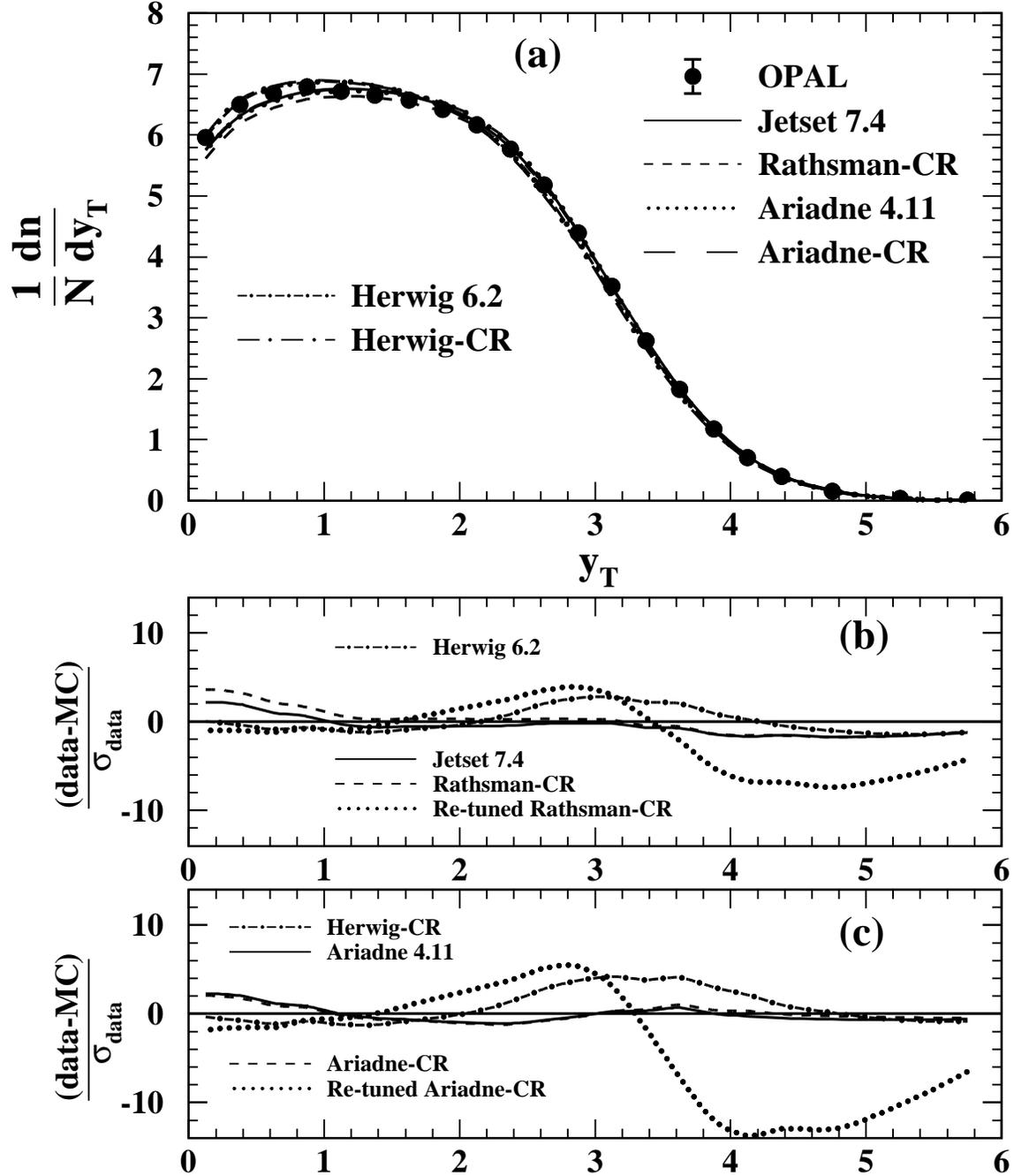}
\end{center}
\caption{
(a) The $y_{T}$ distribution for inclusive Z$^0$ events,
in comparison to
the predictions of models with and without color reconnection (CR).
The data have been corrected for initial-state photon radiation
and detector response.
The uncertainties --~both statistical and total~--
are too small to be visible.
(b)~and~(c) show the deviations of the Monte Carlo predictions
from the data in units of
the total experimental uncertainties, $\sigma_{\mathrm data}$.
}
\label{fig-yt}
\end{figure}

The corrected measurements of
$S$, $A$, $-\ln (y_{34})$ and $y_T$
are presented in Figs.~\ref{fig-sphericity}--\ref{fig-yt}.
These data are consistent with our 
previously published results~\cite{bib-opal1990}.
The data are shown in comparison to 
the predictions of the models at the hadron level.
The model predictions are generally seen to be
similar to each other and in agreement with the experiment.
Parts (b) and (c) of Figs.~\ref{fig-sphericity}--\ref{fig-yt}
show the deviations
of the Monte Carlo predictions from the data in units
of the total experimental uncertainties ``$\sigma_{\mathrm data}$,''
with statistical and systematic terms added in quadrature.
The statistical uncertainties are negligible
compared to the systematic uncertainties.
The curves labelled ``Re-tuned Rathsman-CR''
and ``Re-tuned Ariadne-CR'' in parts (b) and~(c)
are discussed in Sect.~\ref{sec-parameters}.

\begin{table}[t]
\begin{center}
\begin{tabular}{|c|cccc|c|c|}
 \hline
  & & & & & & \\[-2mm]
    Model & $S$  & $A$ & $-\ln (y_{34})$ & $y_{T}$ & Total 
  & $\mnch$ \\
(Number of bins) & (19) & (15) & (26) & (21) & (81) & \\[2mm]
 \hline
  & & & & & & \\[-2mm]
Ariadne     &  1.7 &  6.8 & 10.7 & 17.7 &  36.9 & 21.06 \\
Ariadne-CR  &  6.2 &  5.9 &  4.3 & 16.0 &  32.4 & 21.09 \\
Jetset      & 18.4 & 91.4 & 64.1 & 26.8 & 200.7 & 21.09 \\
Rathsman-CR & 18.5 &103.6 & 74.7 & 46.7 & 243.5 & 20.80 \\
Herwig      & 19.3 & 27.0 & 42.5 & 39.1 & 127.9 & 21.14 \\
Herwig-CR   & 10.5 & 15.7 & 30.6 & 94.8 & 151.6 & 21.06 \\[2mm]
 \hline
  & & & & & & \\[-2mm]
Re-tuned Ariadne-CR
            & 498.1 & 548.7 & 1001.3 & 971.2 & 3019.3 & 21.12 \\[-1mm]
($\ptmin$$\,=\,$$4.7$~GeV/$c,b=0.17$~GeV$^{-2}$)  & & & & & & \\[1mm]
Re-tuned Rathsman-CR
            & 106.8 & 294.2 & 429.7 & 287.0 & 1117.7 & 21.16 \\[-1mm]
($Q_{0}$$\,=\,$$5.5$~GeV/$c^2,b=0.27$~GeV$^{-2}$)  & & & & & & \\[2mm]
 \hline
\end{tabular}
\caption{$\chi^{2}$ values
between the data and models for the distributions shown
in Figs.~\ref{fig-sphericity}--\ref{fig-yt},
calculated using the full experimental uncertainties
including systematic terms.
The number of bins in each distribution is given in 
parentheses in the second row.
The last two rows give the results for re-tuned versions of
the Ariadne-CR and Rathsman-CR models,
see Sect.\ref{sec-parameters}.
The models' predictions for the mean charged particle
multiplicity in inclusive Z$^0$ decays,
$\mnch$, are listed in the last column.
}
\label{tab-chi2}
\end{center}
\end{table} 

We calculated the $\chi^2$ values between the 
hadron level predictions
of the models and the corrected data.
The $\chi^2$ values were determined using 
the total experimental uncertainties,
with no accounting for correlations between the
different bins or distributions.
The $\chi^2$ results are listed in Table~\ref{tab-chi2}.
These $\chi^2$ values are intended to be used only as a
relative measure of the description of the data by the models.
Since the uncertainties are dominated by systematics and
correlations are not considered,
these $\chi^2$ values cannot be used to determine
confidence levels assuming the uncertainties are
distributed according to a normal distribution.
In particular,
a good description does not imply that a model's 
$\chi^2$ should approximately equal the number of data bins.

From Table~\ref{tab-chi2},
it is seen that the $\chi^2$ results for Ariadne 
are much smaller than for Jetset or Herwig.
The reason for this is partly that the detector level
distributions are better described by Ariadne,
as stated above,
and partly that Ariadne is used to determine the
correction factors.
It is unavoidable that
the correction procedure introduces a bias
towards the model used to perform the corrections,
as discussed for example in~\cite{bib-opal1990}.
These biases --~although small~--
can have a significant effect on the $\chi^2$ values
because of the small experimental uncertainties.
For this reason,
it is only meaningful to compare our $\chi^2$ results
within the context of a specific parton shower and
hadronization scheme,
e.g.~Ariadne with Ariadne-CR but not Ariadne with Jetset.

The total $\chi^2$ for the Ariadne-CR model is seen to be
about the same as for Ariadne
(in fact it is a little smaller).
For the Herwig-CR model,
the total $\chi^2$ is about 20\% larger than for Herwig.
From Table~\ref{tab-chi2},
it is seen that this difference
arises entirely from a single distribution,~$y_T$,
however.
Similarly,
the total $\chi^2$ for the Rathsman-CR model
is about 20\% larger than for Jetset,
with the largest contribution to the increase from~$y_T$.

The last column in Table~\ref{tab-chi2} lists the 
predictions of the models for the 
mean value of charged particle multiplicity,
$\mnch$.
The statistical uncertainties are negligible.
The results of all models agree with the 
LEP-averaged result for Z$^0$ decays,
$\mnch$$\,=\,$$21.15\pm 0.29$~\cite{bib-dremingary},
to within the uncertainties.

Thus the three models with color reconnection yield
overall descriptions of the global properties of hadronic
Z$^0$ events which are essentially equivalent to those of
the corresponding models without reconnection.
For the Ariadne-CR and Rathsman-CR models,
this agrees with the
observations in~\cite{bib-lonnblad} and~\cite{bib-rathsman},
respectively.


\section{Gluon jet selection}
\label{sec-selection}

To define jets,
we use the Durham jet finder~\cite{bib-durham}.
The resolution scale, $\ycut$,
is adjusted separately for each event so that exactly three jets
are reconstructed.
The jets are assigned energies using the technique of
calculated energies with massive kinematics,
see for example~\cite{bib-opalgincl98}.
This method relies primarily on the angles between jets 
and the assumption of energy-momentum conservation.
Jet energies determined in this manner are more accurate than
visible jet energies,
with the latter defined by a sum over the reconstructed
energies of the particles assigned to the~jet.

To identify which of the three jets is the gluon jet,
we reconstruct displaced secondary vertices in the quark 
(q or $\mathrm\overline{q}$) jets
and thereby anti-tag the gluon jet.
Displaced secondary vertices are associated with 
heavy quark decay,
especially that of the b quark.
At LEP, b quarks are produced almost exclusively
at the electroweak vertex\footnote{About 22\% of
hadronic Z$^0$ events contain a b$\overline{\mathrm{b}}$
quark pair from the electroweak decay of the Z$^0$,
compared to only about 0.3\%~\cite{bib-lepbb} with
a b$\overline{\mathrm{b}}$ pair from
gluon splitting.}:
thus a jet containing a b hadron is almost always a 
quark jet.
To reconstruct secondary vertices in jets,
we employ the method described in~\cite{bib-qg95a}.
Briefly,
charged tracks are selected for the secondary 
vertex reconstruction procedure if they are assigned
to the jet by the jet finder,
have coordinate
information from at least one of the two silicon detector layers,
a momentum larger than 0.5~GeV/$c$,
and a distance of closest approach
to the primary event vertex~\cite{bib-qg95a} 
less than 0.3~cm.
In addition,
the uncertainty on the distance of closest approach 
must be less than~0.1~cm.
A secondary vertex is fitted using the so-called
``tear down'' method~\cite{bib-qg95a} and is 
required to contain at least three such tracks.
For jets with such a secondary vertex,
the signed decay length, $L$,
is calculated with respect to the primary vertex,
along with its error,~$\sigma_L$.
The sign of $L$ is determined by summing the 3-momenta
of the tracks fitted to the secondary vertex;
$L$$\,>\,$0 if the secondary vertex is displaced from the primary
vertex in the same hemisphere as this momentum sum,
and $L$$\,<\,$0 otherwise.
To be identified as a quark jet,
a jet is required to have a successfully reconstructed secondary vertex
with $L/\sigma_L>2.0$ if it is the highest energy jet or
$L/\sigma_L>5.0$ if it is one of the two lower energy jets.
We require the highest energy jet and exactly
one of the two lower energy jets to be identified as quark jets.
The other lower energy jet is tagged
as a gluon jet.

For each tagged gluon jet,
we determine the scale, $\qjet$,
given by
\begin{equation}
   \qjet = E_{\mathrm jet}\,\sin\left(
         \frac{\theta_{\mathrm min.}}{2} \right)
  \label{eq-scale}
\end{equation}
where $E_{\mathrm jet}$ is the energy of the jet,
with $\theta_{\mathrm min.}$ the smaller of the angles 
between the gluon jet and the other two jets.
Note that due to QCD coherence,
the properties of a gluon jet in {\epem} annihilations 
depend on a transverse momentum-like quantity such as $\qjet$
and not the jet energy,
see for example~\cite{bib-dok88}.
$\qjet$ as defined in eq.~(\ref{eq-scale})
was shown to be an appropriate scale 
for gluon jets in~\cite{bib-kappascale}.

\begin{figure}[t]
\begin{center}
  \epsfxsize=16cm
  \epsffile{pr379_06.epsi}
\end{center}
\caption{
Distribution of the $\qjet$ scale of tagged gluon jets,
see eq.~(\ref{eq-scale}).
The distribution includes the effects of initial-state photon radiation
and detector acceptance and resolution.
The uncertainties are statistical only.
The results are shown in comparison to the predictions of
QCD Monte Carlo programs which include detector simulation
and the same analysis procedures as are applied to the data.
To define hard, acollinear gluon jets,
the region $\qjet$$\,\geq\,$7~GeV,
to the right of the vertical dashed line,
is selected.
The hatched area shows the quark jet background
evaluated using Jetset.
}
\label{fig-qjet}
\end{figure}

The $\qjet$ distribution of the tagged gluon jets
is shown in Fig.~\ref{fig-qjet}.
The data are presented in comparison to the predictions
of the detector level QCD models introduced in
Sect.~\ref{sec-models}.
All the simulations are seen to provide a good
description of the measured $\qjet$ spectrum.

To select hard, acollinear gluon jets,
we require $\qjet$$\,\geq\,$7~GeV.
Further, we require the energy of the gluon jets
to be less than 35~GeV because
the simulations predict the gluon jet
purity (see below) drops sharply for higher energies.
The jets are required to contain at least two particles.
With these cuts, the number of selected gluon jets 
is~$12\,611$.
The energy of the jets varies from about
10~GeV up to the cutoff of 35~GeV,
with an average and RMS of 21.7~GeV and 6.6~GeV,
respectively.

To evaluate the purity of the gluon jets,
we use Monte Carlo samples at the detector level.
We determine the directions of the primary
quark and antiquark from the Z$^0$ 
decay after the parton shower has terminated.
The detector level jet closest to 
the direction of an evolved primary quark or antiquark
is considered to be a quark jet.
The distinct jet closest to the evolved primary
quark or antiquark not associated with this first
jet is considered to be the other quark jet.
The remaining jet is the gluon jet.
The estimated gluon jet purity found using Jetset
is approximately constant at 98\% 
for jet energies from 10 to 25~GeV, 
then decreases to 78\% at 35~GeV.
The overall purity is $(94.6\pm 0.1\,\mathrm{(stat.)})$\%.
Similar results are obtained using all other models
except for Ariadne-CR.\footnote{For the Ariadne-CR model,
the estimated purity is smaller, about 72\%.
Since this model does not describe our gluon jet measurements
well (see Sect.~\ref{sec-cr}),
it is not clear if this estimate is reliable, however.
Note the estimates of gluon jet purity
are presented for informational purposes only.}
Note the overall purity of the gluon jets
decreases to 86\%
after the requirement of a rapidity gap is imposed,
see Sect.~\ref{sec-gap}.
The reason the purity is lower if a rapidity gap 
is required is because gluon jets have a larger mean multiplicity
than quark jets~\cite{bib-gmult},
making it less likely a gap will occur
in gluon jets compared to quark jets as 
the result of a fluctuation.
By requiring the presence of a rapidity gap,
the relative proportion of quark jets is therefore enhanced.

\section{Rapidity gap analysis}
\label{sec-gap}

To identify gluon jets with a rapidity gap,
we examine the charged and neutral particles assigned to 
the selected gluon jets by the jet finder.
The rapidities of the particles are determined
with respect to the jet axis.
The particles in the jet are 
ordered by their rapidity values.

\begin{figure}[tp]
\begin{center}
   \epsfxsize=14cm
   \epsffile{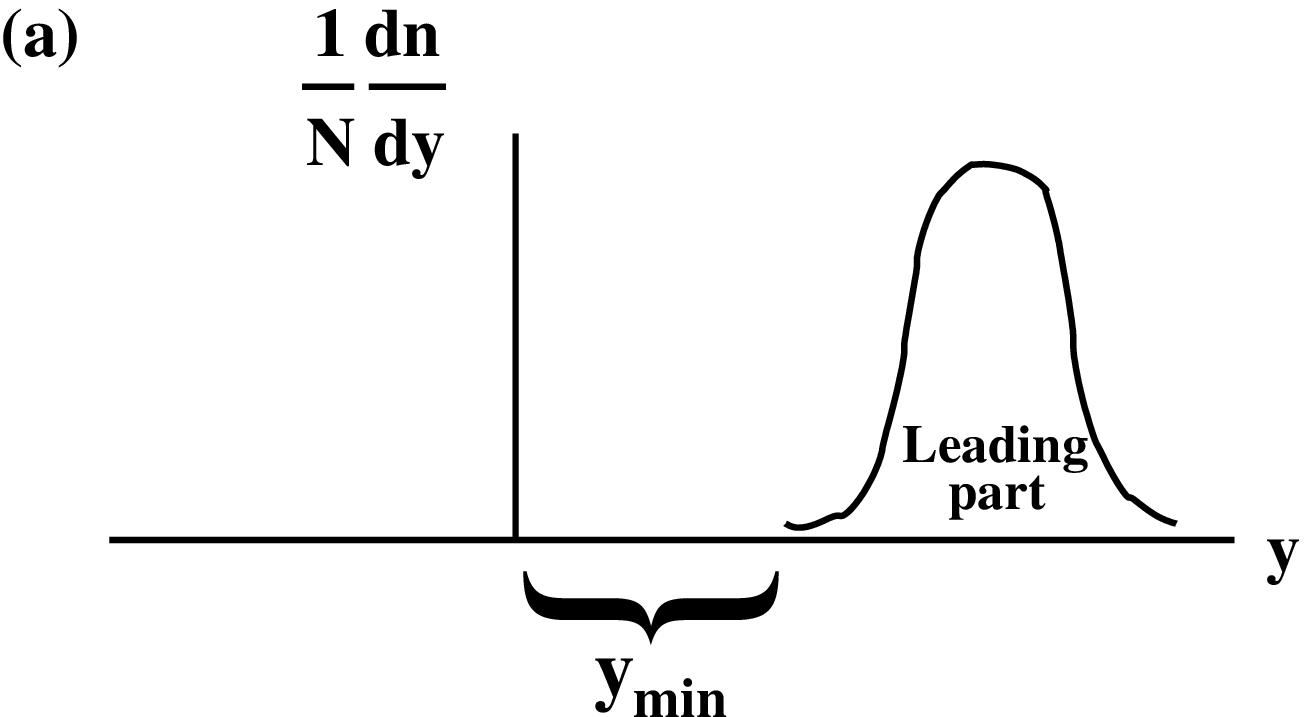} \\[3mm]
   \epsfxsize=14cm
   \epsffile{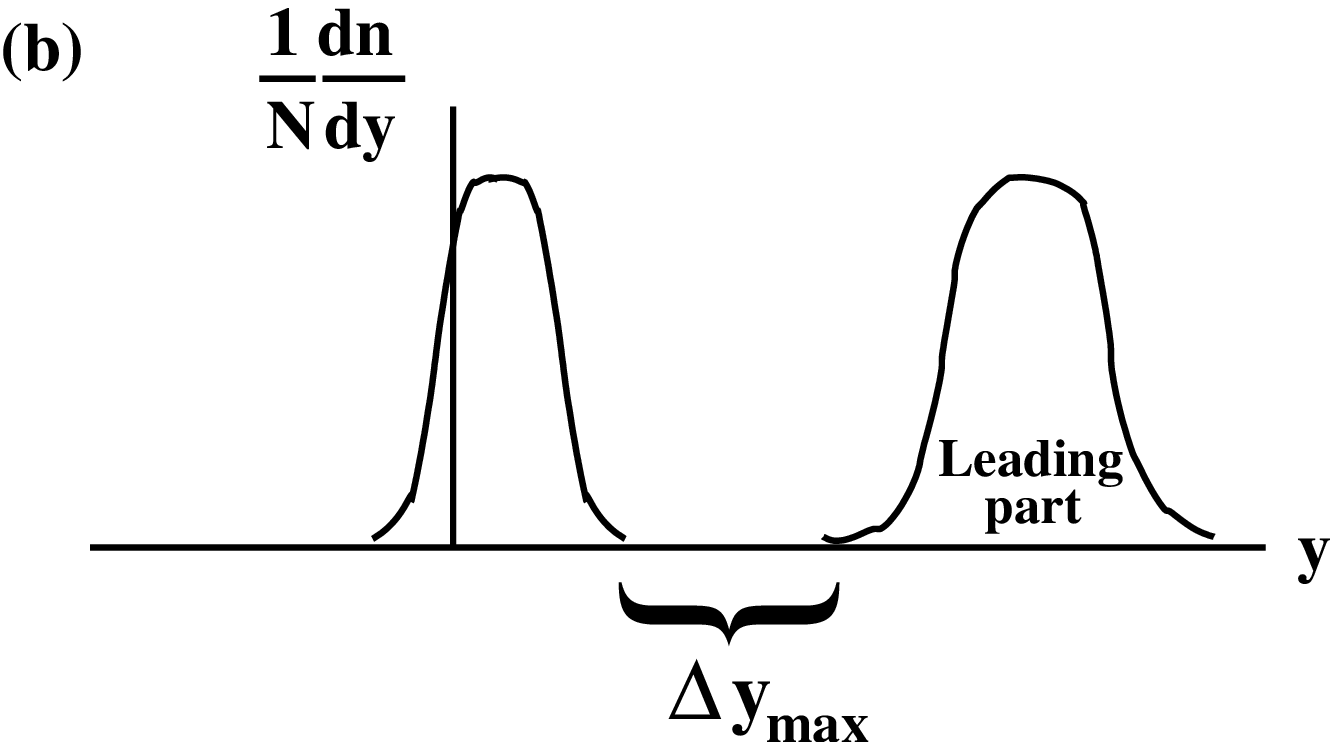}
\end{center}
\caption{Schematic illustration of the distribution 
of particle rapidities
for gluon jets with a rapidity gap as defined in this study:
(a)~for the $\ymin$ sample (see text),
and (b)~for the $\dymax$ sample.
The leading parts of the gluon jets are defined by
charged and neutral particles with rapidities beyond
the gap,
as indicated in the figure.
}
\label{fig-cartoon}
\end{figure}

Models with color reconnection are expected to yield
more events with a large rapidity gap than models
without reconnection,
as discussed in the Introduction.
A large rapidity gap can correspond to a large
value for the smallest particle rapidity
in a jet, $\ymin$,
or else to a large value for the maximum difference between 
the rapidities of adjacent rapidity-ordered particles,
$\dymax$.
These two types of rapidity gap conditions
are illustrated schematically in Fig.~\ref{fig-cartoon}.
Note that the Durham jet finder occasionally assigns particles
to a jet even if the angle between the particle and 
jet axis is greater than 90$^\circ$.
This explains the negative rapidity
values illustrated for
some particles in Fig.~\ref{fig-cartoon}b.

The measured distribution of $\ymin$
is presented in Fig.~\ref{fig-ymin}a.
The data are shown in comparison to the predictions
of the models at the detector level.
To emphasize the difference between models with
and without color reconnection,
we form the following ratio:
\begin{equation}
  \delta_{\ymin} = \frac{f(\ymin)_{\mathrm CR}\, 
      - f(\ymin)_{\mathrm no\, CR} }
   {f(\ymin)_{\mathrm no\, CR}}
 \label{eq-delta}
\end{equation}
with $f(\ymin)_{\mathrm CR}$ the prediction of a model
with color reconnection for a bin of the
$\ymin$ distribution in Fig.~\ref{fig-ymin}a
and $f(\ymin)_{\mathrm no\, CR}$ the prediction 
of the corresponding model without reconnection.
The results for $\delta_{\ymin}$
are shown in Fig.~\ref{fig-ymin}b.
For the Rathsman-CR model,
a significant excess of events
is observed relative to Jetset
for $\ymin$ values larger than about 1.4,
and similarly for the Ariadne-CR model relative to Ariadne.
The Herwig-CR model exhibits a similar excess
with respect to Herwig,
although with less significance.
Based on these results,
we choose $\ymin$$\,\geq\,$1.4 to select 
a sample of gluon jets
with a rapidity gap,
see the dashed vertical line in Fig.~\ref{fig-ymin}b.
In the following,
we refer to this as the ``$\ymin$'' sample.

\begin{figure}[tp]
\begin{center}
  \epsfxsize=15cm
  \epsffile{pr379_08.epsi}
\end{center}
\caption{
(a) Distribution of~$\ymin$ in the tagged gluon jets.
The distribution includes the effects of initial-state photon radiation
and detector acceptance and resolution.
The uncertainties are statistical only.
The results are shown in comparison to the predictions of
QCD Monte Carlo programs which include detector simulation
and the same analysis procedures as are applied to the data.
The hatched area shows the quark jet background
evaluated using Jetset.
(b)~Fractional difference between the results of a
Monte Carlo program with color reconnection and the
corresponding model without reconnection.
To define gluon jets with a rapidity gap,
the region $\ymin$$\,\geq\,$1.4,
to the right of the vertical dashed line,
is selected.
}
\label{fig-ymin}
\end{figure}

\begin{figure}[tp]
\begin{center}
  \epsfxsize=15cm
  \epsffile{pr379_09.epsi}
\end{center}
\caption{
(a) Distribution of~$\dymax$ for tagged gluon jets
with $\ymin$$\,<\,$1.4.
The distribution includes the effects of initial-state photon radiation
and detector acceptance and resolution.
The uncertainties are statistical only.
The results are shown in comparison to the predictions of
QCD Monte Carlo programs which include detector simulation
and the same analysis procedures as are applied to the data.
The hatched area shows the quark jet background
evaluated using Jetset.
(b)~Fractional difference between the results of a
Monte Carlo program with color reconnection and the
corresponding model without reconnection.
To define gluon jets with a rapidity gap,
the region $\dymax$$\,\geq\,$1.3,
to the right of the vertical dashed line,
is selected.
}
\label{fig-dymax}
\end{figure}

For gluon jets with $\ymin$$\,<\,$1.4,
we measure $\dymax$.
The resulting distribution
is shown in Fig.~\ref{fig-dymax}a.
In analogy to eq.~(\ref{eq-delta}),
we form the fractional difference
$\delta_{\dymax}$.
The distribution of $\delta_{\dymax}$
is shown in Fig.~\ref{fig-dymax}b.
A significant excess of events
is observed for the Ariadne-CR and Rathsman-CR models,
relative to Ariadne and Jetset,
for $\dymax$ larger than about 1.3.
We therefore choose $\dymax$$\,\geq\,$1.3 to select
an additional sample of gluon jets
with a rapidity gap,
see the dashed vertical line in Fig.~\ref{fig-dymax}b.
In the following,
we refer to this as the ``$\dymax$'' sample.
For the Herwig-CR model,
there is not a clear excess of events 
relative to Herwig for any $\dymax$ value,
suggesting this distribution is not sensitive
to color reconnection as implemented in Herwig.
In the following,
we therefore test the Herwig-CR model
using the $\ymin$ sample only,
not the standard data set defined by the 
$\ymin$ and $\dymax$ samples taken together.

In total,
655 gluon jets with a rapidity gap are selected,
496  in the $\ymin$ sample
and 159 in the $\dymax$ sample.
The purity of the gluon jets,
evaluated using the method described 
in Sect.~\ref{sec-selection},
is approximately 94\% for gluon jet energies
between 10 and 25~GeV,
then drops to about 50\% at 35~GeV.
The overall purity is 
$(85.7\pm1.0\,({\mathrm stat.}))$\%.
Our subsequent study is based on the leading 
part of these jets,
defined by charged and neutral particles 
with $y$$\,\geq\,$$\ymin$
for events\footnote{For events in this class,
this is therefore the entire jet,
see Fig.~\ref{fig-cartoon}a.} 
in the $\ymin$ sample
and by particles with rapidities beyond
the gap $\dymax$ for events
in the $\dymax$ sample,
see Fig.~\ref{fig-cartoon}.

\section{Color reconnection study}
\label{sec-cr}

To remain as sensitive as possible to color reconnection,
we first compare the Monte Carlo distributions to the data at
the detector level.
Following this,
we correct the measurements for the effects of initial-state radiation,
detector acceptance and resolution,
and gluon jet impurity,
and compare the 
predictions of the models to the data at the hadron level.
The hadron level study allows us to more readily assess the
effect of adjusting Monte Carlo parameters,
see Sect.~\ref{sec-parameters}.

The distributions presented in this section
are normalized to the total number of selected gluon
jets discussed in Sect.~\ref{sec-selection},
i.e. to the number of gluon jets
before the rapidity gap requirement.
The reason for this is to remain sensitive 
to the rate at which gluon jets with a rapidity gap occur,
e.g. to the production rate of events like Fig.\ref{fig-planar}b.

\subsection{Detector level distributions}
\label{sec-uncorrected}

\begin{figure}[tp]
\begin{center}
  \epsfxsize=16cm
  \epsffile{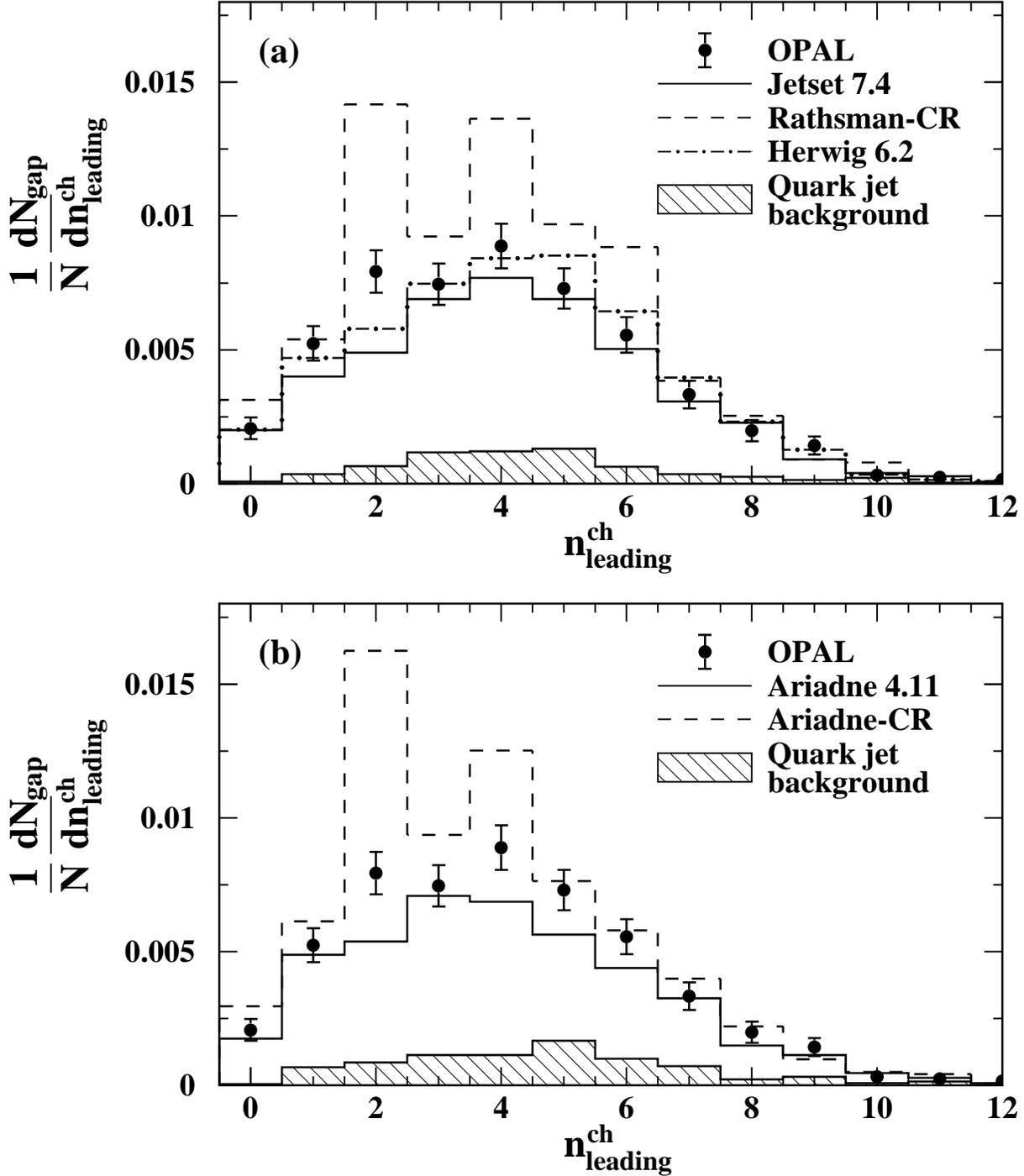}
\end{center}
\caption{
Distribution of $\nchlead$
in the leading part of gluon jets,
based on our standard selection.
``N'' represents the total number of selected gluon jets
and ``N$_{\mathrm gap}$'' the number of gluon jets with
a rapidity gap.
The distribution includes the effects of initial-state photon radiation
and detector acceptance and resolution.
The uncertainties are statistical only.
The results are shown in comparison to the predictions of
QCD Monte Carlo programs which include detector simulation
and the same analysis procedures as are applied to the data:
(a)~the Jetset, Rathsman-CR and Herwig models,
and (b)~the Ariadne and Ariadne-CR models.
The hatched area shows the quark jet background
evaluated using Jetset in part (a) and Ariadne in part~(b).
}
\label{fig-nchlead} 
\end{figure}

The charged particle multiplicity distribution of the 
leading part of the gluon jets, $\nchlead$, 
is shown in Fig.~\ref{fig-nchlead}a.
The results are shown
in comparison to the predictions of the Jetset
and Rathsman-CR models.
Fig.~\ref{fig-nchlead}b shows the same data
compared to Ariadne and \mbox{Ariadne-CR.}
The most striking feature of these results
is the large excess of entries predicted by the
\mbox{Ariadne-CR} and Rathsman-CR models at $\nchlead$$\,=\,$2 and~4
compared to the corresponding models without color reconnection.
Using Monte Carlo information,
we verified these excesses are a consequence 
of events like Fig.~\ref{fig-planar}b,
present in the CR models but not in 
the models without~CR.
The isolated, electrically neutral gluonic system in the 
leading part of the gluon jets in these events
decays into an even number of charged particles,
yielding the spikes at $\nchlead$$\,=\,$2 and~4.

The data are generally well described
by Jetset (Fig.~\ref{fig-nchlead}a),
except for the bins with $\nchlead$$\,=\,$1, 2 and~4
where the data exceed the predictions
by more than one standard deviation of the
statistical uncertainties.
The description by Ariadne 
(Fig.~\ref{fig-nchlead}b)
is considerably worse in that the data lie well 
above the Ariadne results for most of the
range between $\nchlead$$\,=\,$2 and~6.
Nonetheless,
Jetset and Ariadne provide a much better overall
description of the data than the corresponding
models with reconnection.
In particular,
there is not a significant ``spiking effect'' 
in the data at even values of multiplicity
as predicted by these two CR models.
We conclude that color reconnection as implemented
by the \mbox{Rathsman-CR} and Ariadne-CR models is strongly
disfavored,
at least using their standard parameters
given in Sect.~\ref{sec-models}.



\begin{figure}[t]
\begin{center}
  \epsfxsize=16cm
  \epsffile{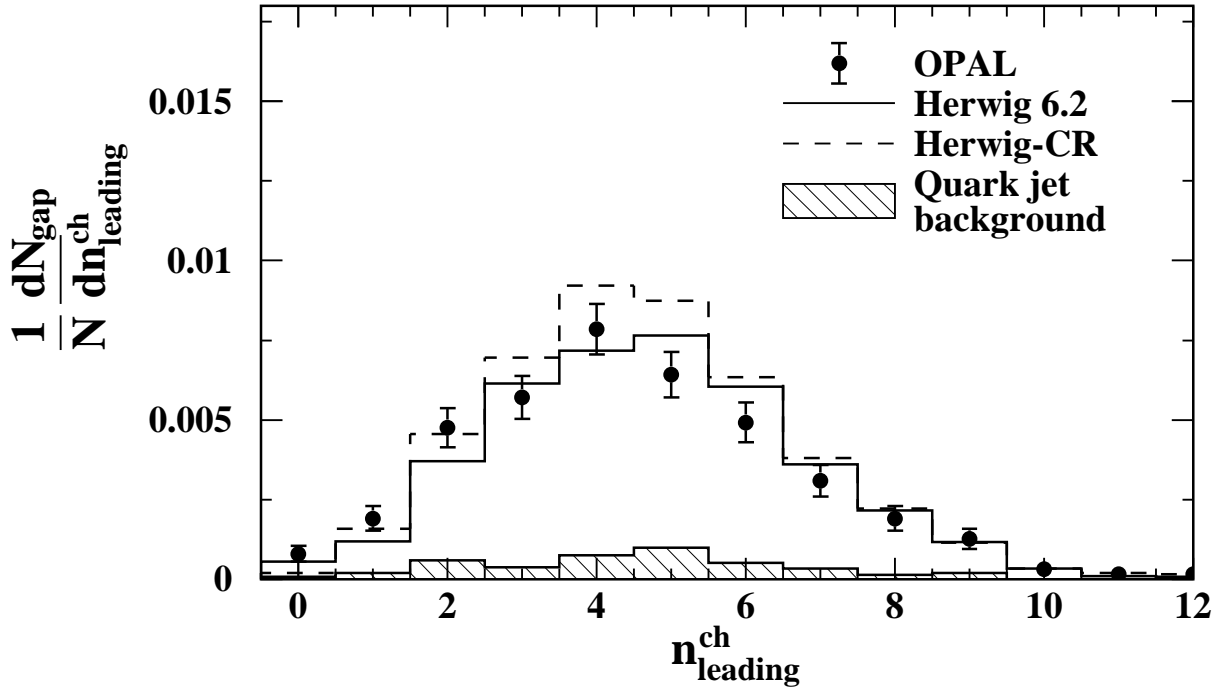}
\end{center}
\caption{
Distribution of $\nchlead$
in the leading part of gluon jets,
based on the $\ymin$ selection
(see Sect.~\ref{sec-gap}).
``N'' represents the total number of selected gluon jets
and ``N$_{\mathrm gap}$'' the number of gluon jets with
a rapidity gap.
The distributions include the effects of initial-state 
photon radiation and detector acceptance and resolution.
The results are shown in comparison to the predictions of
the Herwig and Herwig-CR models.
The uncertainties are statistical only.
The hatched area shows the quark jet background
evaluated using Herwig.
}
\label{fig-nchlead2} 
\end{figure}

The $\nchlead$ distribution obtained using the
$\ymin$ selection 
(see Sect.~\ref{sec-gap})
is presented in Fig.~\ref{fig-nchlead2}.
The data are shown in comparison to the 
corresponding results of the Herwig and Herwig-CR models.
We use the $\ymin$ selection to test Herwig-CR,
and not 
the standard selection defined
by the combined $\ymin$ and $\dymax$ samples,
because the latter is not sensitive to differences
between the Herwig and Herwig-CR models 
as discussed in Sect.~\ref{sec-gap}.
For purposes of comparison,
the prediction of Herwig using the standard
selection is shown in Fig.~\ref{fig-nchlead}a,
however.

From Fig.~\ref{fig-nchlead2},
the Herwig-CR model is seen to predict
a systematic excess of entries relative to 
the corresponding model without~CR
for multiplicities between about 2 and~5.
The overall description of the $\nchlead$ distribution
by the Herwig-CR model is nonetheless reasonable,
at least in comparison to the predictions 
of Jetset and Ariadne in Fig.~\ref{fig-nchlead}.
The best overall description of the $\nchlead$ distribution
is provided by Herwig.

\begin{figure}[tp]
\begin{center}
  \epsfxsize=16cm
  \epsffile{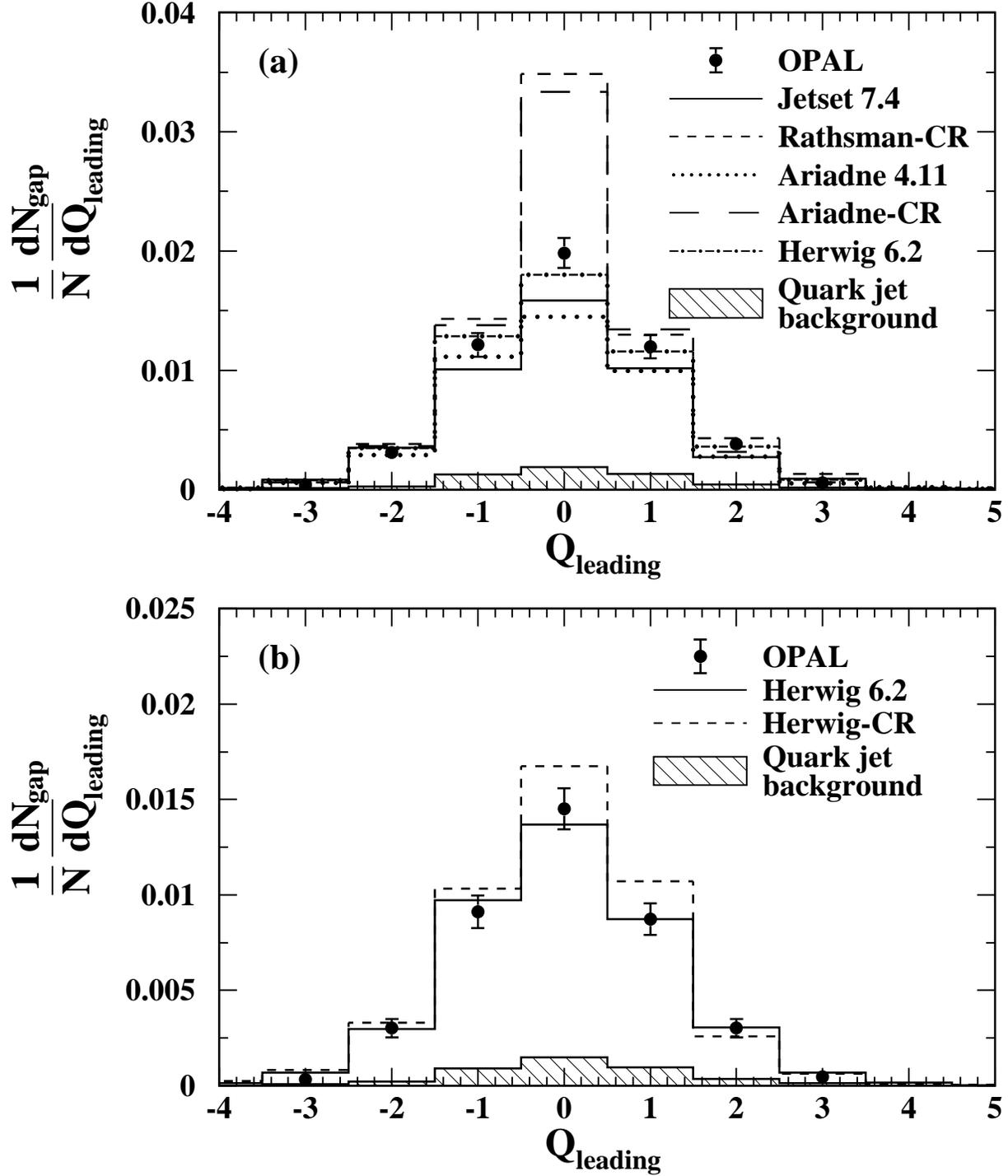}
\end{center}
\caption{
Distribution of $\qtot$
in the leading part of gluon jets in comparison to
the predictions of QCD Model Carlo programs:
(a)~using the standard selection;
(b)~using the $\ymin$ selection.
``N'' represents the total number of selected gluon jets
and ``N$_{\mathrm gap}$'' the number of gluon jets with
a rapidity gap.
For both parts~(a) and~(b),
the distributions include the effects of initial-state photon radiation
and detector acceptance and resolution.
The uncertainties are statistical only.
The hatched area shows the quark jet background
evaluated using Herwig.
}
\label{fig-qtotlead} 
\end{figure}

We next sum the charges of the particles in the leading part
of the gluon jets to find the total leading electric charge,~$\qtot$.
This type of distribution
was suggested in~\cite{bib-ochs}.
The distribution of $\qtot$
is shown in Fig.~\ref{fig-qtotlead}a.
The Rathsman-CR and Ariadne-CR models 
are seen to predict a
large excess of events with $\qtot$$\,=\,$0
compared to the data or models without reconnection,
due to the presence of electrically neutral isolated gluonic systems
at large rapidities as discussed above.
The Jetset and Ariadne predictions for the 
rate of gluon jets with $\qtot$$\,=\,$0
are about 20\% too low.
For purposes of comparison,
the prediction of Herwig is
shown in Fig.~\ref{fig-qtotlead}a.
Herwig is seen to describe the data well.

Our measurement of
the rate of gluon jets with $\qtot$$\,=\,$0
therefore lies between the predictions
of the Jetset and Rathsman-CR models,
and similarly between the predictions of
Ariadne and Ariadne-CR.
In this respect,
the data appear to be consistent with the presence of a
finite amount of color reconnection,
at least as predicted by these two CR models,
albeit at a significantly smaller level than predicted
by the default CR settings of the models.
The most unambiguous signal for color
reconnection in our study is the spiking effect at even
values of $\nchlead$ seen in Fig.~\ref{fig-nchlead}, however.
The data do not provide clear evidence for these spikes.
Furthermore,
the Herwig model without CR describes 
the $\qtot$ distribution well,
as seen from Fig.~\ref{fig-qtotlead}a.
Therefore,
the discrepancies
of Jetset and Ariadne with the data in Fig.~\ref{fig-qtotlead}a
do not provide unambiguous evidence for reconnection effects,
but instead 
are consistent with other inadequacies in the simulations,
not related to CR.
The same statement holds for the discrepancies of
Jetset and Ariadne with the data in Fig.~\ref{fig-nchlead}.

The distribution of $\qtot$ 
obtained using the $\ymin$ selection
is presented in Fig.~\ref{fig-qtotlead}b.
The data are shown in comparison to the corresponding
results from Herwig and \mbox{Herwig-CR.}
Herwig describes the data well,
similar to Fig.~\ref{fig-qtotlead}a.
The predictions of the \mbox{Herwig-CR} model
are seen to lie somewhat above the data,
especially for $\qtot$$\,=\,$0 and~1.

As a systematic check,
we repeated the analysis presented above 
using different choices for the scale of gluon jets,~$\qjet$,
see eq.~(\ref{eq-scale}). 
Specifically we examined the results
for 4$\,<\,$$\qjet$$\,<\,$7~GeV and $\qjet$$\,<\,$4~GeV,
rather than $\qjet$$\,>\,$7~GeV as in our
standard analysis.
Note the definition of a gluon jet becomes
ambiguous for small $\qjet$ values.
We find that the spikes at even values of $\nchlead$
predicted by the Rathsman-CR and Ariadne-CR models
become much less prominent for the smaller $\qjet$ scales,
especially the spike at $\nchlead$$\,=\,$4,
i.e. the selections with softer or
more collinear gluon jets are less
sensitive to color reconnection.
This justifies the choice $\qjet$$\,>\,$7~GeV
for our standard analysis.
To the extent that a CR signal is still visible using the
smaller $\qjet$ ranges,
we find that the values
of $\ymin$ and $\dymax$ above which the predictions of the
CR models exhibit deviations from the
non-CR models are similar to those shown in
Figs.~\ref{fig-ymin}b and~\ref{fig-dymax}b.

As an additional check,
we repeated the analysis described
in Sects.~\ref{sec-selection} and~\ref{sec-gap}
except using energy ordering to identify gluon jets
rather than secondary vertex reconstruction.
In the energy ordering method,
the jet with the smallest calculated energy 
in three-jet $\mathrm q\overline{q}g$ events is
assumed to be the gluon jet.
The purity of gluon jets identified using this technique
is much lower than found using secondary vertices,
especially 
for the high energy jets most sensitive to color reconnection.
The gluon jet purity found using energy ordering is 64\%,
compared to 95\% for our standard analysis.
To increase the purity,
we therefore required  $E_{\mathrm jet}$$\,<\,$15~GeV,
rather than $E_{\mathrm jet}$$\,<\,$35~GeV 
as in the standard analysis.
This method yields about $94\,000$ tagged gluon jets.
The mean gluon jet energy is 12.9~GeV and the estimated purity 81\%.
After imposing the rapidity gap requirements of Sect.~\ref{sec-gap},
we obtain $6604$ gluon jets with an estimated purity of~56\%.
The results we obtain from this check are consistent with our
observations presented above.
In particular,
the results for the 
$\qtot$ distribution are qualitatively similar 
to those shown in Fig.~\ref{fig-qtotlead}.
We note, however,
that the spike at $\nchlead$$\,=\,$4 
predicted by the Rathsman-CR and Ariadne-CR models
in Fig.~\ref{fig-nchlead} is not visible in the corresponding
Monte Carlo predictions based on energy ordering,
because of their softer energy scales
(i.e.~this is similar to the check employing
smaller $\qjet$ values,
mentioned above).
Therefore the selection using energy ordering is not as sensitive
to color reconnection as our standard selection.

The results of Figs.~\ref{fig-nchlead}--\ref{fig-qtotlead}
demonstrate the sensitivity of our study
to processes with color reconnection.
We discuss the effect of parameter variation on the
predictions of the Rathsman-CR and \mbox{Ariadne-CR} models
in Sect.~\ref{sec-parameters}.

\subsection{Correction procedure}
\label{sec-correction}

As the next step in our study,
we correct the data in 
Figs.~\ref{fig-nchlead}--\ref{fig-qtotlead}
to the hadron level.
The correction procedure employs an unfolding matrix.
The matrix is constructed using detector level Monte Carlo events.
The events are subjected to the detector level requirements of
Sects.~\ref{sec-detector} and~\ref{sec-selection}.
In addition,
the events are required to exhibit a rapidity gap,
defined by the conditions of Sect.~\ref{sec-gap},
at both the detector and hadron levels.
The matrices relate the values of $\nchlead$ and $\qtot$ at 
the detector level to the corresponding values
before the same event is processed by the detector simulation.
Thus the matrices correct the data to the hadron level with
the exception that initial-state radiation and the experimental
event acceptance are included.
In a second step,
the data are corrected for event acceptance,
initial-state radiation, and gluon jet impurity
using bin-by-bin factors.
The matrices and bin-by-bin factors are determined
using Herwig because the data in 
Figs.~\ref{fig-nchlead}--\ref{fig-qtotlead}
are best described by that model.
Statistical uncertainties are evaluated for the corrected
data using propagation of errors,
including the statistical uncertainties
of the correction factors.

Because of finite acceptance,
especially for soft particles,
significantly more events satisfy the
rapidity gap requirements at the detector level 
than at the hadron level.
As a consequence,
the overall corrections are fairly large,
of the order of 40\%.
To verify the reliability of the correction procedure,
we therefore performed the following test.
We treated our sample 
of Jetset events at the detector level as ``data,''
using the Herwig derived corrections to unfold them.
The corrected Jetset distributions were found to
agree with the corresponding Jetset distributions
generated at the hadron level to within the statistical
uncertainties.
This demonstrates that our correction procedure does not
introduce a significant bias.

To evaluate systematic uncertainties for the corrected data,
we repeated the analysis using the three systematic variations
listed in Sect.~\ref{sec-inclusive},
with one exception:
to determine the systematic uncertainty related to
the model dependence of the correction factors,
we repeated the analysis using the
Jetset, Ariadne and Herwig-CR models only.
We did not include the Rathsman-CR or Ariadne-CR model
because of their poor description of
the data,
see Figs.~\ref{fig-nchlead} and~\ref{fig-qtotlead}a.
In addition,
we made the following change to the standard analysis
to assess the effect of altering the criteria used
to identify gluon jets.
\begin{itemize}
  \item To identify the lower energy quark jets,
    we required the decay length to satisfy $L/\sigma_L>3$
    rather than $L/\sigma_L>5$;
    this resulted in 1002 tagged gluon jets which satisfied
    the rapidity gap requirements,
    with an estimated purity of 76\%.
\end{itemize}
The systematic uncertainties were treated as described
in Sect.~\ref{sec-inclusive},
i.e. the full differences of the results of the systematic checks
with respect to the standard analysis defined
the systematic uncertainty for each term,
and the individual terms were added in quadrature to
define the total systematic uncertainties.

The largest contributions to the total
systematic uncertainties arose from using
Ariadne to determine the correction factors,
followed by the requirement $L/\sigma_L>3$
to identify the lower energy quark jets.

\subsection{Hadron level distributions}
\label{sec-hadron}

\begin{figure}[tp]
\begin{center}
  \epsfxsize=16cm
  \epsffile{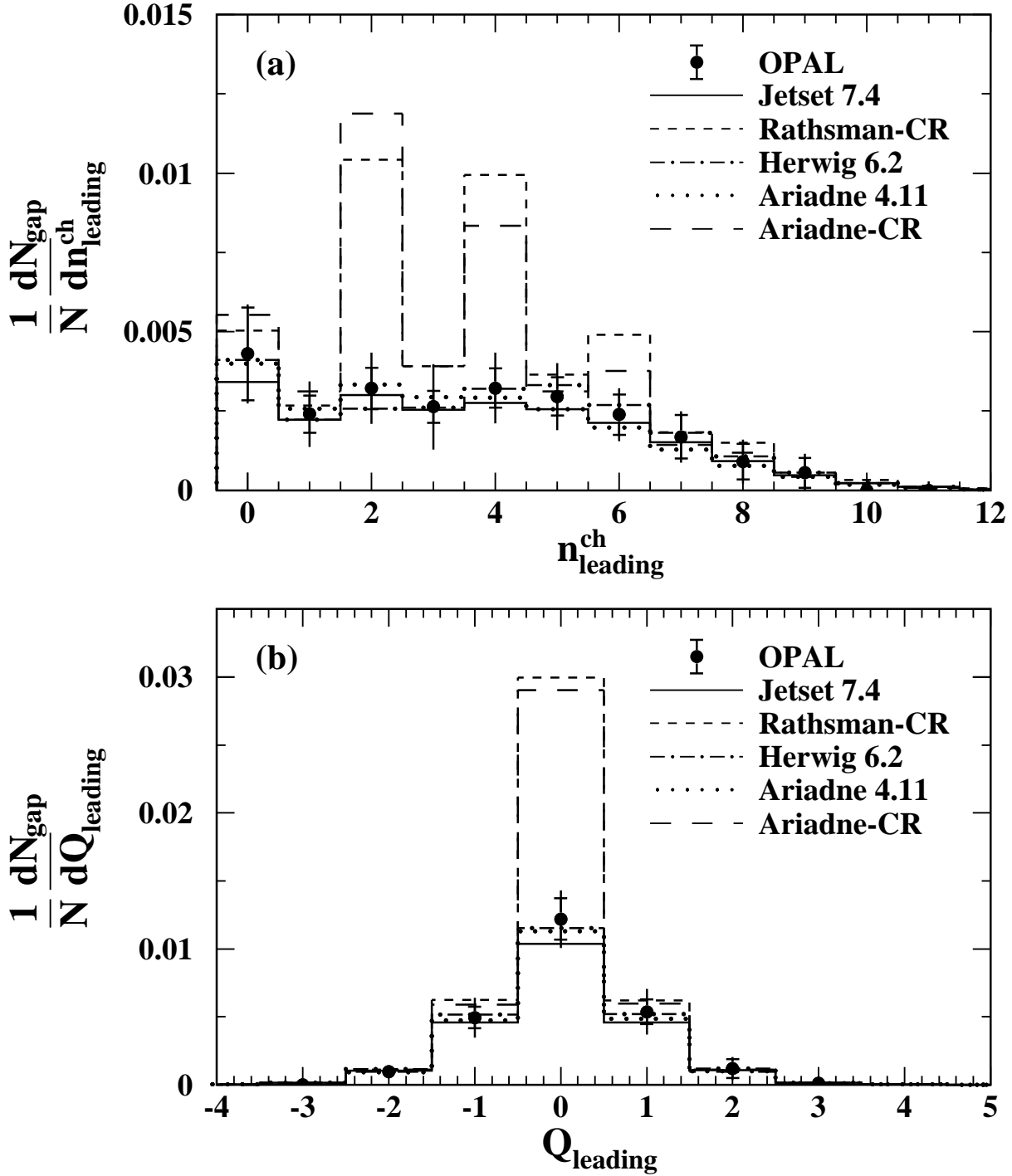}
\end{center}
\caption{
Distributions of (a)~$\nchlead$ and (b)~$\qtot$
in the leading part of gluon jets,
based on our standard selection.
``N'' represents the total number of selected gluon jets
and ``N$_{\mathrm gap}$'' the number of gluon jets with
a rapidity gap.
The data have been corrected for initial-state photon radiation,
gluon jet impurity,
and detector response.
The horizontal bars indicate the statistical uncertainties.
The vertical lines show the total uncertainties,
with statistical and systematic terms added in quadrature.
The results are shown in comparison to the predictions of
QCD Monte Carlo models.
}
\label{fig-nchleadcorr}
\end{figure}

The corrected distributions of
$\nchlead$ and $\qtot$
are presented in Fig.~\ref{fig-nchleadcorr}.
These results are based on our standard selection,
i.e.~the $\ymin$ and $\dymax$ samples added together.
The data are shown in comparison to the hadron level 
predictions of the Jetset, Rathsman-CR, 
Ariadne and Ariadne-CR models.
For purposes of comparison,
the predictions of Herwig are shown as well.
The qualitative features of the predictions 
are seen to be similar to those of the corresponding 
detector level distributions in 
Figs.~\ref{fig-nchlead} and~\ref{fig-qtotlead}a.
In particular,
the Ariadne-CR and \mbox{Rathsman-CR} models exhibit
a large excess of entries at $\nchlead$$\,=\,$2 and~4
in Fig.~\ref{fig-nchleadcorr}a,
corresponding to the $\qtot$$\,=\,$0 bin in Fig.~\ref{fig-nchleadcorr}b,
analogous to the results of Sect.~\ref{sec-uncorrected}.
From Fig.~\ref{fig-nchleadcorr}a
it is also seen that the Rathsman-CR model
predicts a spike at $\nchlead$$\,=\,$6.
This latter feature was not apparent in the detector
level distribution of Fig.~\ref{fig-nchlead}a
because of finite detector resolution.



The corresponding results based on the $\ymin$ sample
are presented in Fig.~\ref{fig-nchleadcorrhw}.
The data are shown in comparison to the predictions of the
Herwig and Herwig-CR models.
From Fig.~\ref{fig-nchleadcorrhw}a it is seen that the
Herwig-CR model predicts a significant excess of 
events relative to Herwig 
for $\nchlead$$\,=\,$2, 4 and~6,
analogous to the results of the Rathsman-CR 
and Ariadne-CR models in Fig.~\ref{fig-nchleadcorr}a.
This suggests that
the production of events like Fig.~\ref{fig-planar}b
is a general feature of color reconnection.
The spike in the prediction of the
Herwig-CR model at $\nchlead$=4 in Fig.~\ref{fig-nchleadcorrhw}a
probably explains the general excess of the Herwig-CR results
above Herwig
for multiplicities between $\nchlead$=3 and~5
in Fig.~\ref{fig-nchlead2}.


\begin{figure}[tp]
\begin{center}
  \epsfxsize=16cm
  \epsffile{pr379_14.epsi}
\end{center}
\caption{
Distributions of (a)~$\nchlead$ and (b)~$\qtot$
in the leading part of gluon jets,
based on the $\ymin$ selection.
``N'' represents the total number of selected gluon jets
and ``N$_{\mathrm gap}$'' the number of gluon jets with
a rapidity gap.
The data have been corrected for initial-state photon radiation,
gluon jet impurity,
and detector response.
The horizontal bars indicate the statistical uncertainties.
The vertical lines show the total uncertainties,
with statistical and systematic terms added in quadrature.
The results are shown in comparison to the predictions of
the Herwig and Herwig-CR Monte Carlo models.
}
\label{fig-nchleadcorrhw}
\end{figure}

\subsection{Effect of parameter variation on the model predictions}
\label{sec-parameters}

We next study the effect of parameter variation
on the predictions of the Rathsman-CR and
Ariadne-CR models,
to determine if they can be tuned 
to describe the gluon jet data of Fig.~\ref{fig-nchleadcorr}
without adversely affecting their descriptions of
the inclusive Z$^0$ decay measurements
presented in Sect.~\ref{sec-inclusive}.

To begin,
we define $\delqtot$
to be the difference between the Monte Carlo prediction
and experimental result for the
$\qtot$$\,=\,$0 bin in Fig.~\ref{fig-nchleadcorr}b.
We then vary the principal parameters of
the models one at a time,
with the other parameters at their standard values,
to see if it is possible to reduce $\delqtot$ to zero
or near-zero,
i.e.~to obtain agreement of the model's prediction
with this measurement.
We note that if $\delqtot$ is near-zero,
the predictions of the model for the $\nchlead$ distribution in
Fig.~\ref{fig-nchleadcorr}a
will also be in general agreement with the data
since the events which yield the excess of entries
in Fig.~\ref{fig-nchleadcorr}b
are the same as those which yield the excess in
Fig.~\ref{fig-nchleadcorr}a.

\paragraph{The Rathsman-CR model:}
\label{sec-rathsman}

For the Rathsman-CR model,
the following parameters were varied:
\begin{itemize}
  \item $\lamqcd$, the QCD scale parameter,
    given by PARJ(81);
  \item $Q_0$, the minimum mass value to which
    partons evolve, 
    given by PARJ(82);
  \item $a$ and $b$, which control the longitudinal
    momentum spectrum of hadrons relative to the
    string direction in the Lund model of hadronization,
    given by PARJ(41) and PARJ(42);
  \item $\sigma_q$, which controls the transverse
    momentum spectrum of hadrons,
    given by PARJ(21).
\end{itemize}
The PARJ references are the names of the
parameters in the Pythia Monte Carlo.
These five parameters are the most important ones
controlling the multiplicity and momentum distributions
of hadrons in the model.
Note we do not include the color reconnection suppression
factor $R_0$ mentioned in Sect.~\ref{sec-models} in the 
above list.
It is a trivial result that the Rathsman-CR model will
describe the data as well as Jetset for
$R_0$$\,\rightarrow\,$0
since the two models are identical in this limit.
Varying $R_0$ to reproduce the experimental result and
corresponding uncertainty for
the $\qtot$$\,=\,$0 bin in Fig.~\ref{fig-nchleadcorr}b
yields 
$R_0$$\,=\,$$0.0085\pm 0.0075\, {\mathrm (stat.)}\pm 
0.0087\, {\mathrm (syst.)}$,
consistent with $R_0$$\,=\,$0.

\begin{figure}[p]
\begin{center}
  \epsfxsize=17cm
  \epsffile{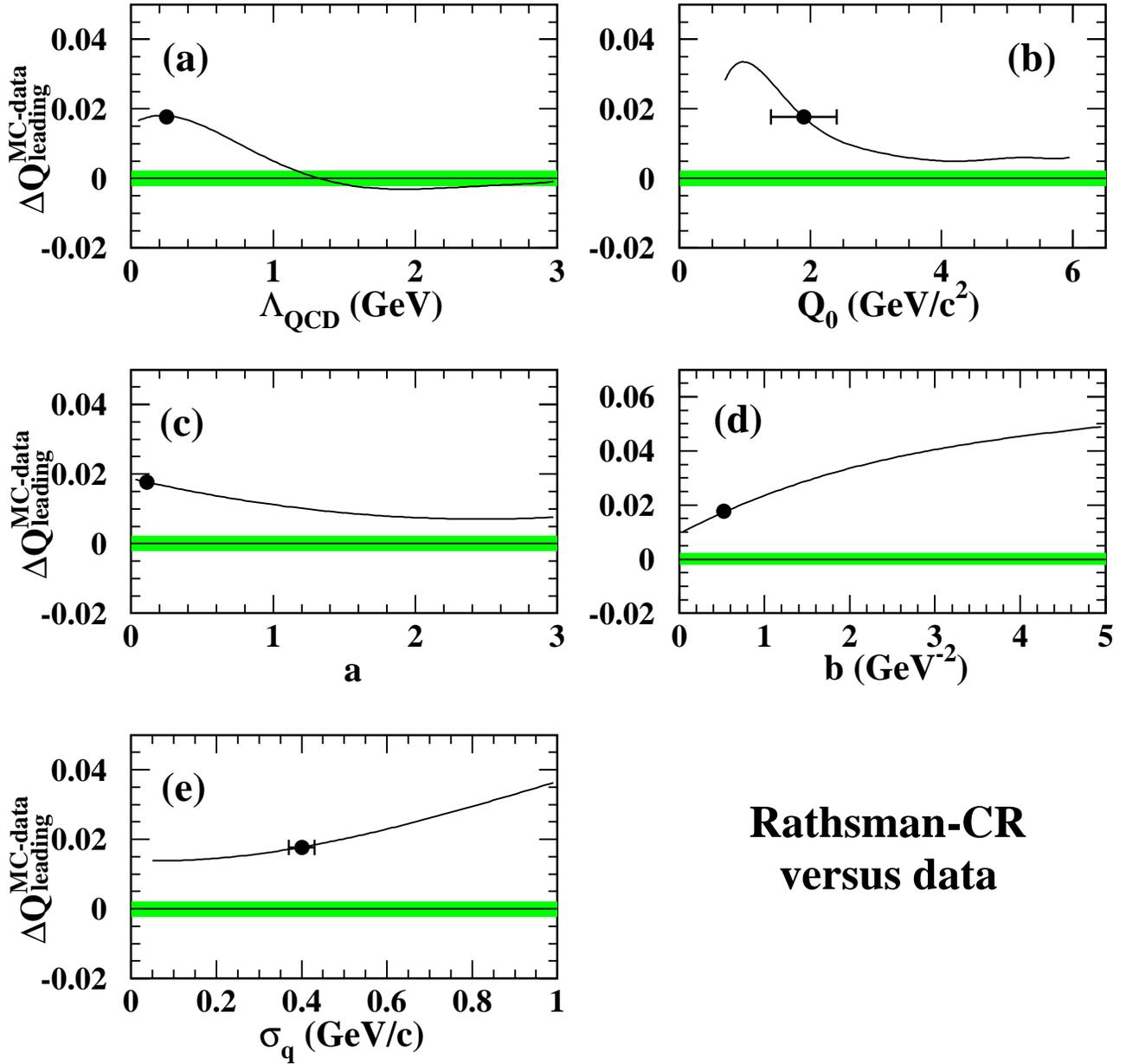}
\end{center}
\caption{
Results for the difference between the 
Rathsman-CR Monte Carlo prediction and the
experimental result for the
$\qtot$$\,=\,$0 bin in Fig.~\ref{fig-nchleadcorr}b,
$\delqtot$,
as the principal parameters of the model
are changed with the other parameters at their
default settings.
The solid dots indicate the standard values
of the parameters.
The horizontal error ranges show the uncertainties
attributed to the parameter values in~\cite{bib-qg95b}.
For the parameters $\lamqcd$ and $b$,
these uncertainties are too small to be visible.
Note an uncertainty is not evaluated for the
$a$ parameter.
The width of the shaded band centered on
$\delqtot$$\,=\,$0 equals twice the total
experimental uncertainty of $\delqtot$.
}
\label{fig-rathsman-vary}
\end{figure}

The results for $\delqtot$
are shown in Fig.~\ref{fig-rathsman-vary}.
The standard values of the parameters are indicated
by solid dots.
The uncertainties attributed to the parameter values
in~\cite{bib-qg95b},
beyond which the description of inclusive Z$^0$ measurements
is significantly degraded if the other parameters 
remain at their standard values,
are indicated by the horizontal error ranges.
Note that an uncertainty is not evaluated for the $a$
parameter in~\cite{bib-qg95b} and that the uncertainties
attributed to $\lamqcd$ and $b$
are too small to be visible.
The width of the shaded bands in Fig.~\ref{fig-rathsman-vary}
indicates twice the total experimental uncertainty of $\delqtot$,
corresponding to plus and minus one standard deviation.

It is seen that $\delqtot$ can be reduced to zero
for $\lamqcd$$\,\approx\,$1.3~GeV.
As $\lamqcd$ is increased,
more soft gluons are produced,
increasing the probability for multiple
color reconnections in an event.
In events with multiple reconnections,
color strings can reconnect the isolated gluonic
string segment illustrated in Fig.~\ref{fig-planar}b
back with the rest of the event,
spoiling the rapidity gap.
From Fig.~\ref{fig-rathsman-vary}
it is also seen that $\delqtot$
can be reduced to near-zero for large values of $Q_0$,
e.g.~$Q_0$$\,\gsim\,$4~GeV/$c^2$.
As $Q_0$ is increased,
fewer soft gluons are available,
effectively decreasing the reconnection probability.
In this sense,
an increase in the value of $Q_0$ is analogous 
to a reduction in the value of the
parameter $R_0$ discussed above.
We note that the values of $\lamqcd$ and $Q_0$
required to reduce $\delqtot$ to zero or near-zero
represent large excursions from their standard values.
Fig.~\ref{fig-rathsman-vary} suggests it is unlikely that
$\delqtot$ can be reduced to zero or near-zero through 
variation of $a$ or~$\sigma_q$. 

Setting $Q_0$ to 3.5~GeV/$c^2$
with the other parameters at their standard values,
the Rathsman-CR model predicts 
a mean charged multiplicity in inclusive
Z$^0$ events of $\mnch$$\,=\,$20.2,
smaller than the experimental result of
$21.15\pm 0.29$ mentioned in Sect.~\ref{sec-inclusive}.
Mean multiplicity in the Lund hadronization model
is primarily controlled by the
parameters $a$ and~$b$.
Therefore,
having set $Q_0$$\,=\,$3.5~GeV/$c^2$,
we varied the $b$ parameter\footnote{The $a$ and $b$
parameters are highly correlated with respect to the
model predictions for $\mnch$;
therefore we consider variation of the $b$ parameter alone,
not both $a$ and~$b$.}
to reproduce the measured
result for $\mnch$.
To increase the prediction for $\mnch$,
$b$ needs to be decreased.
As $b$ decreases, $\delqtot$ also tends to become smaller
(see Fig.~\ref{fig-rathsman-vary}d).
By iterating the adjustment of $Q_0$ and $b$,
it therefore proved possible to simultaneously obtain
$\delqtot$$\,\approx\,$0 and $\mnch$$\,\approx\,$21.15.
The result we find for the two parameters is
$Q_0$$\,=\,$$5.5$~GeV/$c^2$ and $b$$\,=\,$$0.27$~GeV$^{-2}$,
corresponding to $\delqtot$$\,=\,$$6.7\times 10^{-7}$.
We refer to the Rathsman-CR model with these adjusted
parameters as the ``re-tuned'' Rathsman-CR model.

A one standard deviation limit 
was evaluated for the re-tuned parameters
by adjusting $Q_0$ and $b$ so they yielded the correct
result for $\mnch$ and agreement with the one standard
deviation upper limit for $\delqtot$
shown in Fig.~\ref{fig-rathsman-vary}:
the result is $Q_0$$\,=\,$3.7~GeV/$c^2$ 
and $b$$\,=\,$0.35~GeV$^{-2}$.
A two standard deviation limit was evaluated 
in an analogous manner,
based on twice the total uncertainty of $\delqtot$:
the result is $Q_0$$\,=\,$3.2~GeV/$c^2$
and $b$$\,=\,$0.38~GeV$^{-2}$.
Finally,
$Q_0$ and $b$ were adjusted to yield 
\mbox{$\delqtot$$\,\approx\,$0} and a value for
$\mnch$ equal to the LEP-averaged result
plus its one standard deviation uncertainty,
specifically $\mnch$$\,=\,$21.44,
see above and Sect.~\ref{sec-inclusive}.
The result is $Q_0$$\,=\,$5.2~GeV/$c^2$ 
and $b$$\,=\,$0.25~GeV$^{-2}$.

We examined the description of the re-tuned
Rathsman-CR model for the 
inclusive Z$^0$ measurements in 
Figs.~\ref{fig-sphericity}--\ref{fig-yt}.
The total $\chi^2$ for the 81 bins of data 
was found to be~1117.7,
much larger than the result $\chi^2$$\,=\,$243.5 presented 
in Sect.~\ref{sec-inclusive} for the standard version
of the Rathsman-CR model (see Table~\ref{tab-chi2}).
Using the one and two standard deviation limits
for the re-tuned parameters, 
given above,
the corresponding $\chi^2$ are 435.1 and 327.2
respectively,
still significantly larger than the $\chi^2$ of the standard version.
The $\chi^2$ result for the parameters tuned to yield
$\mnch$$\,=\,$21.44 is 785.2.

The $\chi^2$ results for the re-tuned Rathsman-CR model
are listed in the bottom portion of Table~\ref{tab-chi2}.
The deviations of the re-tuned model from
the measured distributions
are shown by the dotted curves in part~(b)
of Figs.~\ref{fig-sphericity}--\ref{fig-yt}.

We attempted to follow an analogous procedure
to that described above for $Q_0$
to adjust~$\lamqcd$.
With $\lamqcd$ set to 1.3~GeV and the
other parameters at their standard values,
the mean charged multiplicity of inclusive Z$^0$ events is~26.4.
To reduce this to 21.15,
we increased the value of~$b$.
As $b$ increases,
$\delqtot$ becomes larger, however
(Fig.~\ref{fig-rathsman-vary}d),
and we could not find a solution which yielded both
$\delqtot$$\,\approx\,$0 and $\mnch$$\,\approx\,$21.15.
The closest solution we found,
defined by the set of parameters 
which provided the correct inclusive multiplicity
and a minimal result for $\delqtot$,
was $\lamqcd$$\,=\,$1.3~GeV and $b$$\,=\,$4.9~GeV$^{-2}$,
which yielded $\delqtot$$\,=\,$0.012,
with $\chi^2$$\,=\,$$2.1\times 10^4$ for the data of
Figs.~\ref{fig-sphericity}--\ref{fig-yt}.

Last,
motivated by the observation that larger values of
$\lamqcd$ and $Q_0$ both reduce $\delqtot$
while having opposite effects on $\mnch$,
we increased both $\lamqcd$ and $Q_0$,
with the other parameters at their standard values,
to search for a solution with
$\delqtot$$\,\approx\,$0 and $\mnch$$\,\approx\,$21.15.
Specifically,
we systematically increased the value of $\lamqcd$
and then performed a scan to determine the value of $Q_0$
which yielded the correct result for $\mnch$.
The parameter set with the minimal result for $\delqtot$
was $\lamqcd$$\,=\,$0.6~GeV and $Q_0$$\,=\,$4.9~GeV/$c^2$,
which yielded $\delqtot$$\,=\,$0.004,
with $\chi^2$$\,=\,$1606 for the data of
\mbox{Figs.~\ref{fig-sphericity}--\ref{fig-yt}.}

We conclude it is unlikely that the gluon jet
results of Sect.~\ref{sec-hadron}
can be reproduced by the Rathsman-CR model
through variation of~$\lamqcd$,
similar to our observation above
for $a$ and $\sigma_q$.

Thus,
the only mechanism we found to adjust the parameters
of the Rathsman-CR model to simultaneously describe
our data on rapidity gaps in gluon jets and
$\mnch$ in inclusive Z$^0$ decays
was to increase $Q_0$ to 
values in the range from about 3.3 to 5.5~GeV/$c^2$,
much larger than the values of 1--2~GeV/$c^2$
normally attributed to this parameter.
These large values of $Q_0$ resulted in
a significant degradation of the model's
description of inclusive Z$^0$ events, however,
as discussed above.
We conclude it is unlikely that this model
can simultaneously provide a satisfactory description
of the data in Sects.~\ref{sec-inclusive}
and~\ref{sec-hadron}
using its standard value for the
strength of color reconnection, 
$R_0$$\,=\,$0.1.
Thus,
our results provide compelling evidence to disfavor
color reconnection as it is currently implemented
by this model.

\paragraph{The Ariadne-CR model:}

The following parameters of the Ariadne-CR model 
were varied to determine their influence on $\delqtot$:
\begin{itemize}
  \item $\lamqcd$, given by PARA(1);
  \item $\ptmin$, the minimum transverse momentum
    of a gluon with respect to the dipole which emits it,
    given by PARA(3);
  \item $a$, $b$ and $\sigma_q$,
    given by PARJ(41), PARJ(42) and PARJ(21)
    as for the Rathsman-CR model.
\end{itemize}
The PARA references are the names of the
parameters in Ariadne,
see Table~\ref{tab-ariadne}.
Analogous to our treatment of the Rathsman-CR model,
we do not include the color suppression factor 
PARA(26) (see Sect.~\ref{sec-models}) in this list.
Varying PARA(26) to reproduce the result
for the $\qtot$$\,=\,$0 bin in Fig.~\ref{fig-nchleadcorr}b,
as well as the result for the $\qtot$$\,=\,$0 bin
plus its one standard deviation total uncertainty,
yields PARA(26)$\,=\,$96 and 41, respectively,
much larger than the standard value PARA(26)$\,=\,$9.
Note that large values of PARA(26) correspond to the limit
of large $N_{\mathrm C}$
in which the
probability for color reconnection becomes negligible.

\begin{figure}[p]
\begin{center}
  \epsfxsize=17cm
  \epsffile{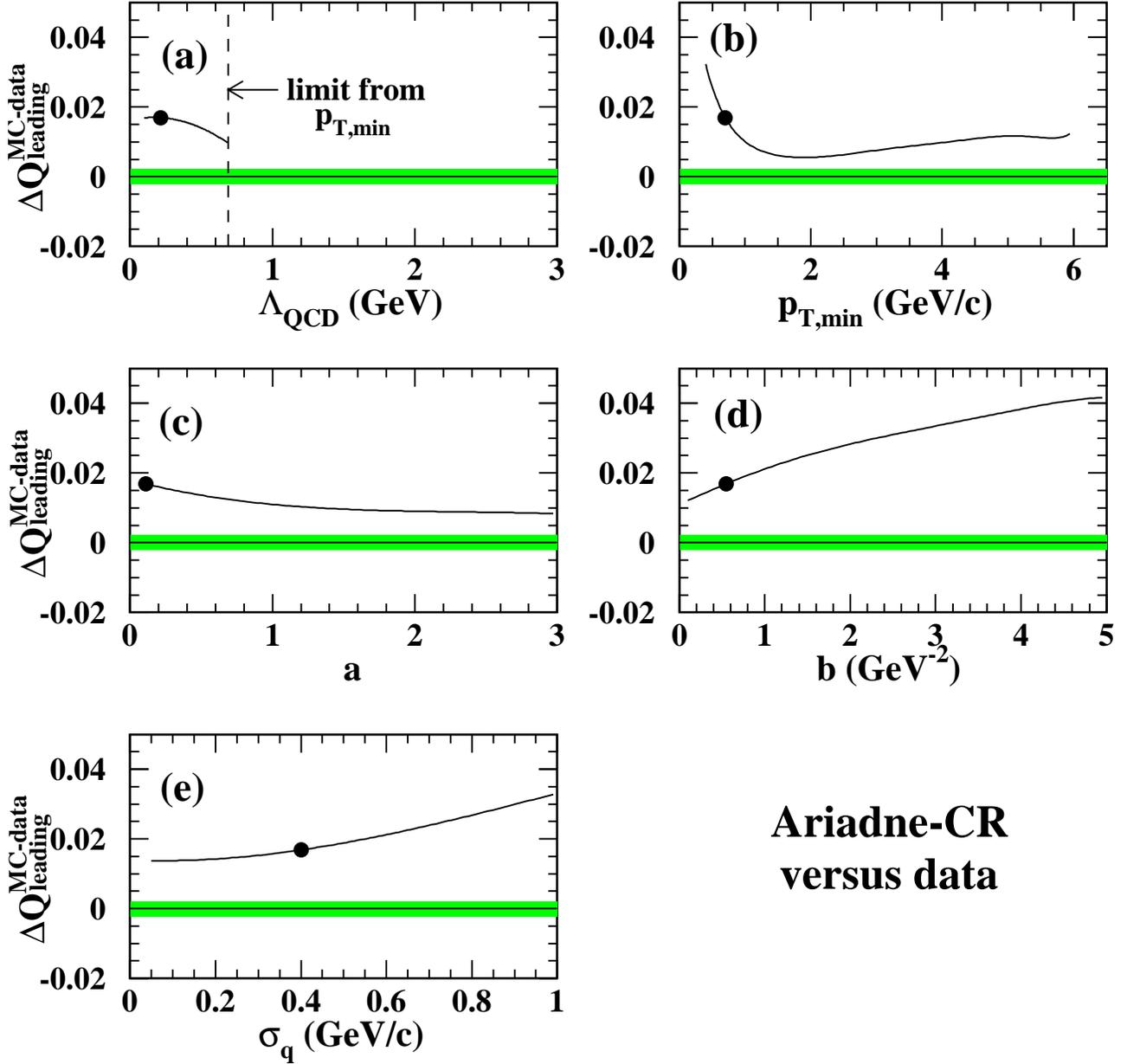}
\end{center}
\caption{
Results for the difference between the 
Ariadne-CR Monte Carlo prediction and the
experimental result for the
$\qtot$$\,=\,$0 bin in Fig.~\ref{fig-nchleadcorr}b,
$\delqtot$,
as the principal parameters of the model
are changed with the other parameters at their
default settings.
The solid dots indicate the standard values
of the parameters.
The uncertainties attributed to the parameter values
in Table~\ref{tab-ariadne} are too small to be visible.
Note an uncertainty is not evaluated for the $a$ 
or $\sigma_q$ parameters.
The width of the shaded band centered on
$\delqtot$$\,=\,$0 equals twice the total
experimental uncertainty of $\delqtot$.
}
\label{fig-ariadne-vary}
\end{figure}

The results are shown in Fig.~\ref{fig-ariadne-vary}.
The standard parameter values are indicated by solid dots.
Their uncertainties as given in Table~\ref{tab-ariadne}
are too small to be visible in the figure.
Note an uncertainty was not evaluated for the $a$ or $\sigma_q$
parameters.
Also note the Ariadne computer program requires
$\lamqcd$$\,<\,$$\ptmin$.
For this reason,
the results for $\lamqcd$ are shown up to 0.70~GeV only,
which is the standard value of $\ptmin$.

The results of Fig.~\ref{fig-ariadne-vary}
are similar to those of Fig.~\ref{fig-rathsman-vary},
i.e.~$\delqtot$ approaches zero as the parton shower
cutoff $\ptmin$ is increased from its standard value,
while it exhibits the same behavior shown
in Fig.~\ref{fig-rathsman-vary} as
$a$, $b$ and $\sigma_q$ are varied.

Setting $\ptmin$$\,=\,$2~GeV/$c$ so that 
$\delqtot$$\,\approx\,$0
(see Fig.~\ref{fig-ariadne-vary}b),
with the other parameters at their standard values,
$\mnch$ in inclusive Z$^0$ decays is predicted to be~20.0.
Through iterative adjustment of $\ptmin$ and $b$,
we found $\delqtot$$\,\approx\,$0
(specifically,
$\delqtot$$\,=\,$$-4.1\times 10^{-5}$)
and $\mnch$$\,\approx\,$21.15
for $\ptmin$$\,=\,$4.7~GeV/$c$ and $b$$\,=\,$0.17~GeV$^{-2}$.
We refer to the Ariadne-CR model with these adjusted
parameters as the ``re-tuned'' Ariadne-CR model.
One and two standard deviation limits
were evaluated for the re-tuned parameters
in the same manner as described above for the re-tuned
Rathsman-CR model;
the results are 
$\ptmin$$\,=\,$2.0~GeV/$c$ and $b$$\,=\,$0.35~GeV$^{-2}$,
and $\ptmin$$\,=\,$1.5~GeV/$c$ and $b$$\,=\,$0.42~GeV$^{-2}$,
respectively.
Similarly,
we tuned $\ptmin$ and $b$ to yield
$\delqtot$$\,\approx\,$0 and $\mnch$$\,=\,$21.44,
analogous to the procedure in Sect.~\ref{sec-rathsman}:
this yielded
$\ptmin$$\,=\,$3.5~GeV/$c$ and $b$$\,=\,$0.20~GeV$^{-2}$.

Using the re-tuned parameters to determine the predictions
of the model for the data in 
Figs.~\ref{fig-sphericity}--\ref{fig-yt}
resulted in a total $\chi^2$ of 3019.3 for those distributions,
compared to the result $\chi^2$$\,=\,$32.4
presented in Sect.~\ref{sec-inclusive} for
the standard version of Ariadne-CR (Table~\ref{tab-chi2}).
For the one and two standard deviation re-tuned parameters,
the corresponding results are $\chi^2$$\,=\,$333.1
and 132.6, respectively.
Thus the description provided by the re-tuned model is much
worse than that provided by the standard version,
even considering the uncertainties of the
re-tuned parameter set.
For the parameters tuned to yield $\mnch$$\,=\,$21.44,
the corresponding $\chi^2$ is~1254.6.

The $\chi^2$ results for the re-tuned Ariadne-CR model
are listed in the bottom portion of Table~\ref{tab-chi2}.
The deviations of the re-tuned model from
the measured distributions
are shown by the dotted curves in
part~(c) of Figs.~\ref{fig-sphericity}-\ref{fig-yt}.

Analogous to the procedure we followed for the Rathsman-CR model,
we also attempted to simultaneously describe our gluon jet
data and $\mnch$ in inclusive Z$^0$ decays
by varying both $\lamqcd$ and $\ptmin$
with the other parameters at their standard values.
The solution which yielded
the minimal result for $\delqtot$
while correctly describing $\mnch$
was $\lamqcd$$\,=\,$0.5~GeV and $\ptmin$$\,=\,$2.4~GeV/$c$,
which yielded $\delqtot$$\,=\,$0.0041,
with $\chi^2$$\,=\,$2095 for the data of
\mbox{Figs.~\ref{fig-sphericity}--\ref{fig-yt}.}
We did not attempt an adjustment of the two parameters
$\lamqcd$ and $b$ as we did for the Rathsman-CR model
since the value of $\lamqcd$ 
is constrained by $\ptmin$ as explained above.

Thus the only manner we found to adjust the parameters
of the Ariadne-CR model to describe our data on gluon jets
with a rapidity gap and at the same time yield
$\mnch$$\,\approx\,$21.15
was to increase the parton shower cutoff parameter $\ptmin$ 
to values in the range from about 1.6 to 4.7~GeV/$c$,
significantly larger than the range of 0.6--0.7~GeV/$c$ normally
attributed to this parameter.
This resulted in a significant degradation of the model's 
description of the global properties of inclusive
Z$^0$ events as described above.
Analogous to the Rathsman-CR model,
we conclude it is unlikely that the Ariadne-CR model can
simultaneously provide a satisfactory description of
the data in both
Sects.~\ref{sec-inclusive} and~\ref{sec-hadron}
using its standard value for the strength of color reconnection,
and that our results provide compelling evidence to
disfavor this model.

\section{Search for glueball-like resonances}
\label{sec-glue}

\begin{figure}[t]
 \begin{center}
 \begin{tabular}{ccc}
      \epsfxsize=4.7cm
      \epsffile{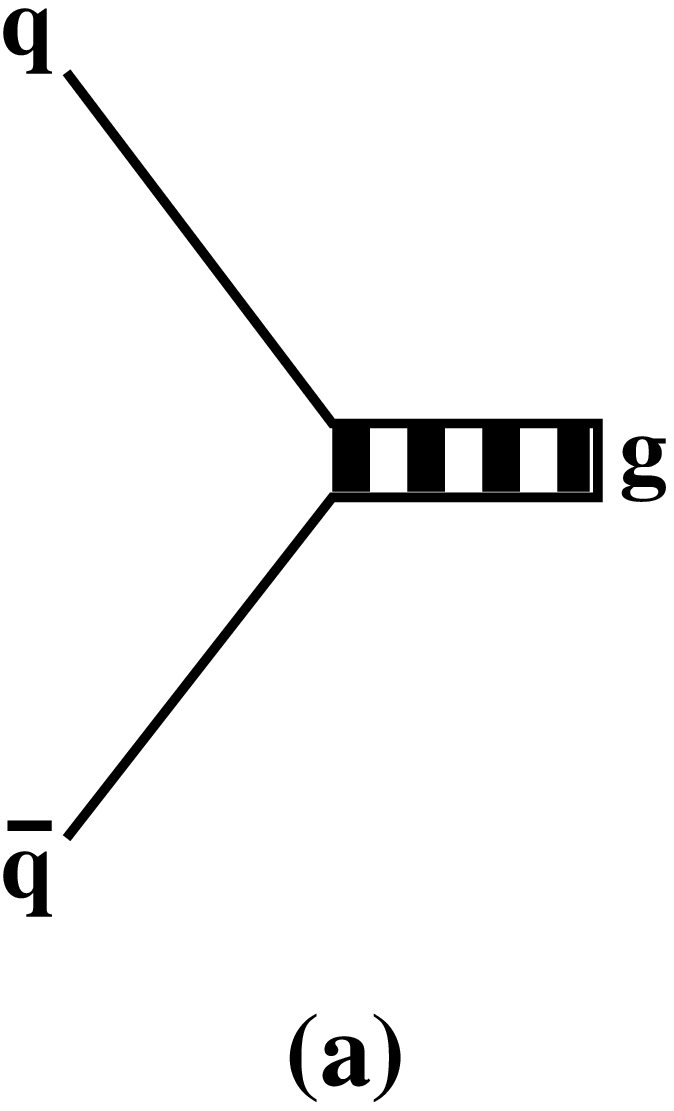} & \hspace*{.25cm}
      \epsfxsize=4.7cm
      \epsffile{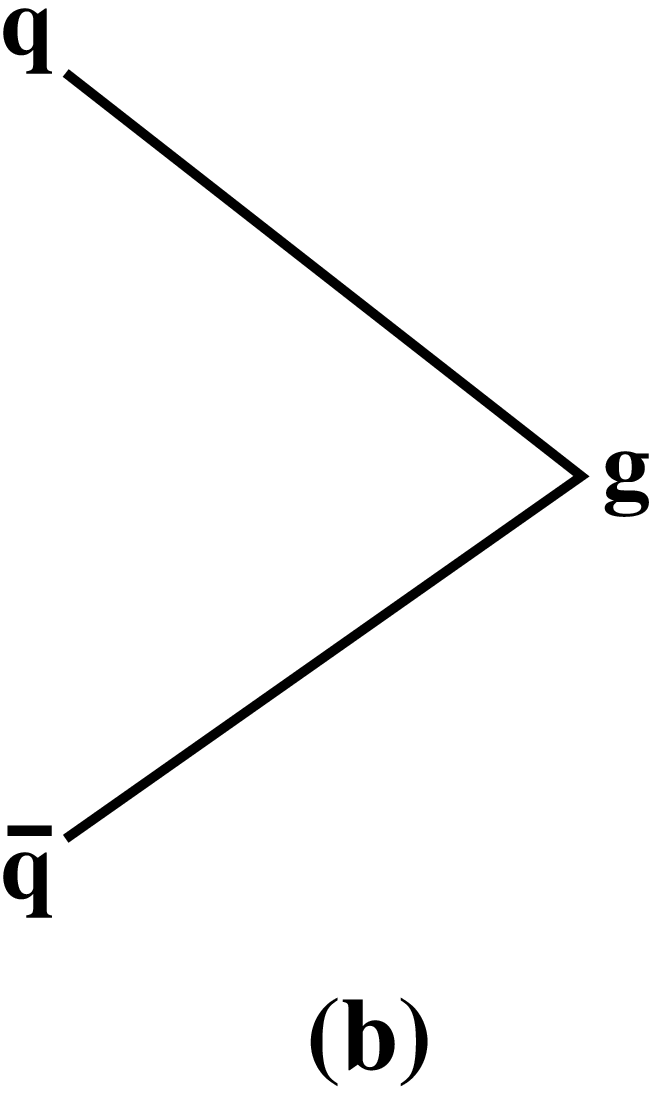}& \hspace*{.25cm}
      \epsfxsize=5.7cm
      \epsffile{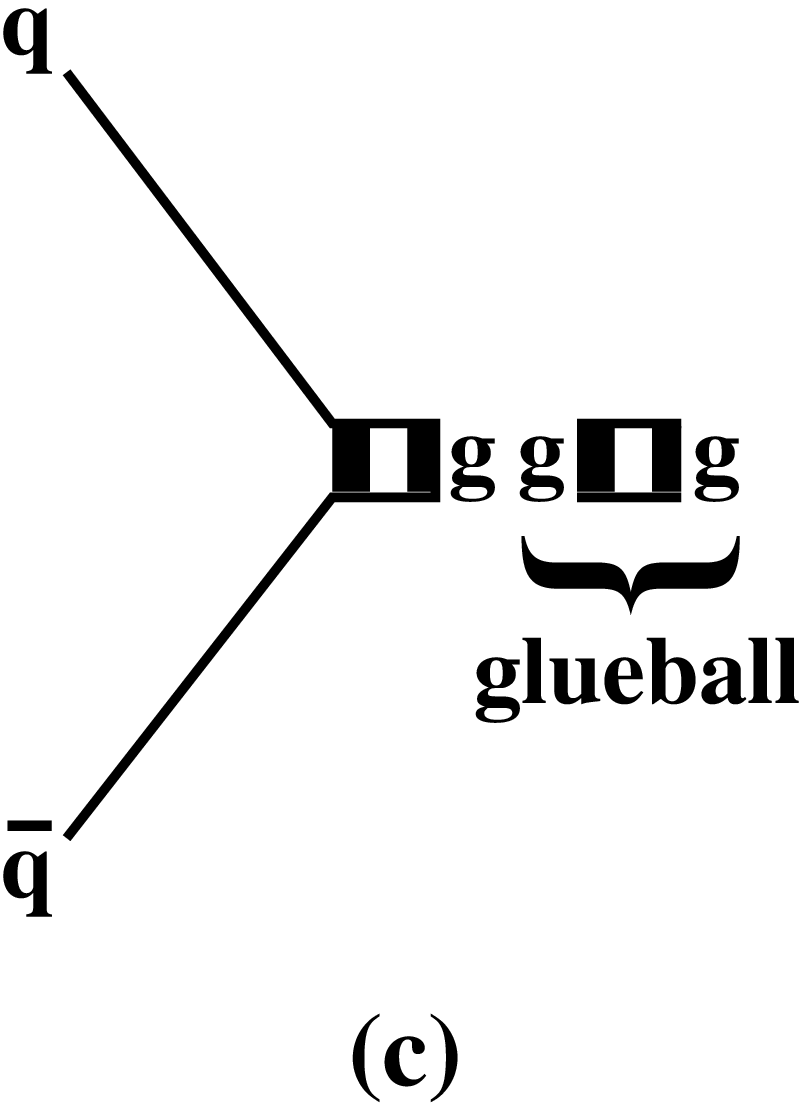}
 \end{tabular}
 \end{center}
\caption{
(a)~Schematic illustration of a three-jet 
$\mathrm q\overline{q}g$ event
with a color octet field stretched between the
gluon g and residual quark-antiquark 
$\mathrm q\overline{q}$ system.
The double line with hatching,
attached to the gluon,
represents the octet field.
The single lines connecting the octet field
to the quark and antiquark represent 
color triplet fields.
(b)~Illustration of a $\mathrm q\overline{q}g$
event in which color triplet fields connect the
gluon directly with the quark and antiquark.
(c)~The octet field can be neutralized by the production
of virtual gg color singlets from the vacuum,
leading to the formation of glueballs.
}
\label{fig-coloroctet}
\end{figure}

Besides providing a sensitive means to
test models of color reconnection,
gluon jets with a rapidity gap present an environment
which may favor the production of glueballs,
as discussed in~\cite{bib-ochs}.
If a hard, acollinear gluon in a {\epem} three-jet 
$\mathrm q\overline{q}g$ event
propagates a significant distance without radiating,
a rapidity gap can form between the
gluon jet and the rest of the event.
This could enhance the probability for a color octet field
to be created between the
gluon and residual $\mathrm q\overline{q}$ system,
see Fig.~\ref{fig-coloroctet}a.
This octet field is analogous to the field which is
expected to connect 
two separating gluons produced in a color singlet state.
This is in contrast to e.g. the Lund hadronization model,
in which only color triplet fields are present,
see Fig.~\ref{fig-coloroctet}b.
Color octet fields provide a natural environment
in which to create glueballs~\cite{bib-peterson},
through gg pair production from the vacuum,
see Fig.~\ref{fig-coloroctet}c.

QCD lattice calculations and other sources
(see~\cite{bib-pdg} and references therein)
suggest that the mass of the lightest glueball state
should lie in the general range from about
1 to 2~GeV/$c^2$.
One of the main candidates for the lightest glueball
is the $f_0(1500)$~\cite{bib-amsler},
with a resonance width of 0.11~GeV/$c^2$.
The principal charged particle decay modes of 
the $f_0(1500)$ are $\pi^+\pi^-$ and $\pi^+\pi^-\pi^+\pi^-$.

Following the suggestions in~\cite{bib-ochs},
we therefore examine invariant mass spectra in the
leading part of our selected gluon jets.
Since we are searching for anomalous resonant structure,
the data are compared to the predictions of
the models without color reconnection,
i.e. Jetset, Ariadne and Herwig.
These models do not contain glueballs.
The distributions are examined at the detector level only
and are normalized to the number of entries in
the distributions.
The bin widths are adjusted to reflect the
mass resolution of the detector,
estimated from the simulations.

\begin{figure}[tp]
\begin{center}
  \epsfxsize=16cm
  \epsffile{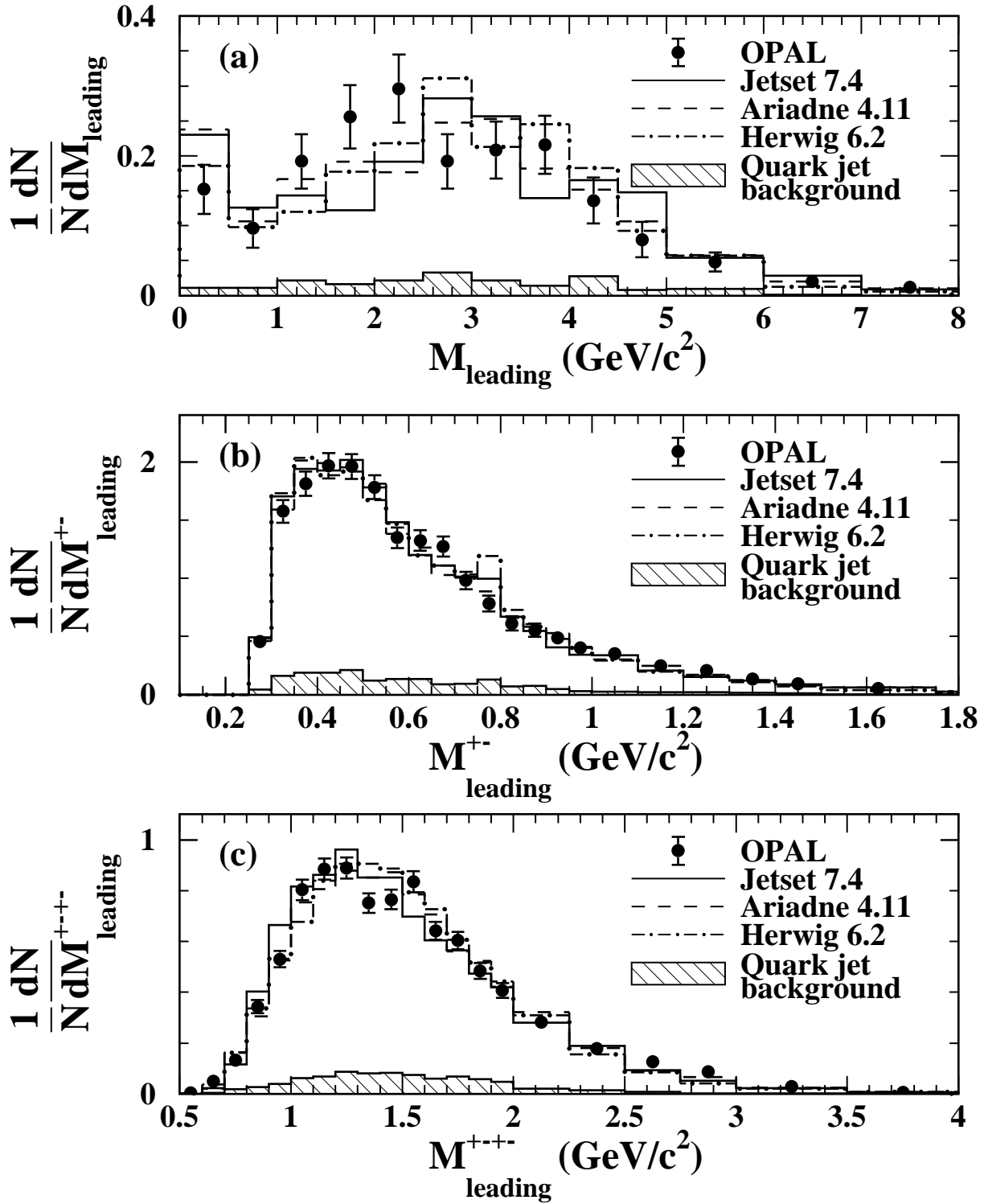}
\end{center}
\caption{
Distribution of jet invariant mass based on charged 
and neutral particles in the leading part of gluon jets.
The distribution includes the effects of initial-state photon radiation
and detector acceptance and resolution.
The uncertainties are statistical only.
The results are shown in comparison to the predictions of
QCD Monte Carlo programs which include detector simulation
and the same analysis procedures as are applied to the data.
The hatched area shows the quark jet background
evaluated using Herwig.
}
\label{fig-mlead}
\end{figure}

We begin by examining the total invariant mass of the 
leading part of the gluon jets, $\mlead$.
These masses are determined
using both charged and neutral particles.
Charged particles are assumed to be pions and neutral particles photons,
as stated in Sect.~\ref{sec-detector}.
Since glueballs are electrically neutral,
we select gluon jets with $\qtot$$\,=\,$0,
see Fig.~\ref{fig-qtotlead}a.
Of the sample of 655 events discussed in Sect.~\ref{sec-gap},
this yields 250 gluon jets.
The $\mlead$ distribution of these jets
is presented in Fig.~\ref{fig-mlead}a.
The data are shown
in comparison to the corresponding results 
of Jetset, Ariadne and Herwig.
The models are seen to 
describe the data reasonably well.
There is a general excess of the data
above the Monte Carlo predictions for three bins
in the mass range from 1.0 to 2.5~GeV/$c^2$.
The total $\chi^2$ values with respect to the data 
for these three bins are 11.5 for Jetset,
7.3 for Herwig and 6.0 for Ariadne.
Since this excess is only about two standard deviations
of the statistical uncertainties above the
predictions of Herwig and Ariadne\footnote{The Monte Carlo 
results are based on about
twice as many events as the experimental distributions,
see Sects.~\ref{sec-detector} and~\ref{sec-models}.},
it is not possible to obtain a definite conclusion
concerning this discrepancy.

Motivated by the charged particle 
decay modes of the $f_0(1500)$, mentioned above,
we also examine the distributions of invariant mass
of two oppositely charged particles
in the leading part of the gluon jets,
$\twomlead$,
and the corresponding distribution of four charged particles 
with total electric charge zero,
$\fourmlead$.
The simulations predict that about 75\% of the charged
particles in the leading part of the selected gluon jets are pions.
The distributions of $\twomlead$ and $\fourmlead$
are presented in Figs.~\ref{fig-mlead}b  and~c.
Since in this case the glueball candidate does not
necessarily comprise the entire leading part of the gluon jet,
there is no reason to constrain $\qtot$ to be zero.
Therefore,
the $\twomlead$ and $\fourmlead$ distributions
are based
on the entire sample of 655 gluon jets with a rapidity gap,
not just the jets with $\qtot$$\,=\,$0.
Again,
the simulations are seen to describe the data 
reasonably well.
The most significant excess of data above the
Monte Carlo predictions
occurs in the tail of the $\fourmlead$ distribution,
at mass values between 2.5 and 3.0~GeV/$c^2$.
This excess is about two standard deviations
of the statistical uncertainties above the
predictions of Ariadne and somewhat larger for
the other models.
Therefore,
we do not observe clear evidence for
anomalous resonant structure.
Note the spike in the Herwig prediction for
$\twomlead$$\,\approx\,\,$0.77~GeV/$c^2$ in Fig.~\ref{fig-mlead}b 
is due to the $\rho$ meson resonance
which is too narrow in Herwig.
Jetset exhibits a similar effect but
at a less significant level.

To better isolate a signal from a scalar particle such as
the $f_0(1500)$,
we also examined the $\cos\theta^*$ distribution of charged
particle pairs in the leading part of the gluon jets,
defined as follows.
A ``parent'' momentum is defined by summing the momenta
of two oppositely charged particles.
$\theta^*$ is the angle between the 
either of the two ``decay'' particles and the parent momentum,
in the rest frame of the parent.
The distribution of $\cos\theta^*$ should be flat for a scalar
particle but not necessarily for the combinatoric background.
The measured $\cos\theta^*$ distribution was found to be
well described by the Monte Carlo simulations.
Therefore,
we do not obtain any evidence
for the anomalous production of scalar particles.

\section{Summary and conclusion}

A sample of $12\,611$
gluon jets with a mean energy of 22~GeV
and estimated purity of 95\% is identified 
in {\epem} hadronic Z$^0$ decay events
using b quark jet tagging.
The data were collected with the OPAL detector at LEP.
A subsample of about 5\% of these jets is selected
which exhibit a rapidity gap,
i.e. an absence of charged and neutral particles
over a significant range of rapidity
as illustrated in Fig.~\ref{fig-cartoon}.
After imposing the rapidity gap requirement,
the estimated purity of the gluon jets is~86\%.

We examine the predictions of three models 
of color reconnection (CR):
the L\"{o}nnblad model~\cite{bib-lonnblad}
(see also~\cite{bib-gostacr})
implemented in the Ariadne Monte Carlo,
the Rathsman model~\cite{bib-rathsman}
implemented in the Pythia Monte Carlo,
and the color reconnection model in the
Herwig Monte Carlo~\cite{bib-herwig2}.
We refer to these as the Ariadne-CR,
Rathsman-CR, and Herwig-CR models, respectively.
Specifically,
we examine the predictions of these models for
the distributions of
charged particle multiplicity and total electric charge
in the leading part of the gluon jets,
defined by charged and neutral particles beyond the gap.

We find that the Rathsman-CR and Ariadne-CR
models predict a large excess of gluon jets with a rapidity gap,
for which the leading part of the jets
is electrically neutral,
compared to the corresponding models without color reconnection.
In particular,
these two models predict large spikes in the
charged particle multiplicity distribution at even
values of multiplicity.
Thus our analysis is very sensitive to CR effects.
We adjust the principal parameters of the two models
to determine if they can be tuned to simultaneously
provide a good description of our gluon jet measurements
and the global properties of inclusive events in 
hadronic Z$^0$ decays.
We find we can obtain a satisfactory description of
the gluon jet data and the mean charged particle
multiplicity $\mnch$ in inclusive Z$^0$ events
only for very large values of the parton shower
cutoff parameters,
$Q_0$$\,\approx\,$3.2--5.5~GeV/$c^2$ for the Rathsman-CR model
or $\ptmin$$\,\approx\,$1.5--4.7~GeV/$c$ for the Ariadne-CR model,
and that the overall description of
global distributions in inclusive Z$^0$ events
is then severely degraded.
We conclude that it seems unlikely that either 
of these two models can be tuned to
provide a satisfactory description of both our gluon jet data and the
global properties of Z$^0$ events, 
using their standard values for the
strength of color reconnection. 
We therefore conclude that color reconnection 
as currently implemented in these models
is disfavored.
Our conclusion for the Ariadne-CR model is consistent
with our previous results~\cite{bib-opalgincl98}.
Here,
we present an even more sensitive study of
color reconnection and systematically examine the effects of
parameter variation on the model's predictions.

The Herwig-CR model also predicts 
a significant excess of events 
at even values of charged particle 
multiplicity in the leading part of gluon jets,
compared to the corresponding model without color reconnection,
cf.~Fig.~\ref{fig-nchleadcorrhw}a.
These excesses are much less prominent than for the
Rathsman-CR and Ariadne-CR models, however,
and are not clearly visible once the effects of
finite detector resolution are incorporated,
cf.~Fig.~\ref{fig-nchlead2}.
Therefore,
we are unable to obtain a definite conclusion
concerning this model.
The data are nonetheless better described by
the version of Herwig without color reconnection.

Our study is also potentially sensitive to the
presence of color singlet, electrically neutral objects
such as glueballs,
see~\cite{bib-ochs}.
We therefore examine the total invariant mass distribution
of the leading part of gluon jets,
using events in which the leading system is electrically neutral.
We also examine the invariant mass distributions of
two oppositely charged particles,
and of four charged particles 
with total electric charge zero,
in the leading part of our sample of gluon jets.
We do not observe any evidence for anomalous features
in the data,
including the production of glueball-like objects.

\section{Acknowledgments}

We particularly wish to thank the SL Division for the efficient operation
of the LEP accelerator at all energies
 and for their close cooperation with
our experimental group.  In addition to the support staff at our own
institutions we are pleased to acknowledge the  \\[2mm]
Department of Energy, USA, \\
National Science Foundation, USA, \\
Particle Physics and Astronomy Research Council, UK, \\
Natural Sciences and Engineering Research Council, Canada, \\
Israel Science Foundation, administered by the Israel
Academy of Science and Humanities, \\
Benoziyo Center for High Energy Physics,\\
Japanese Ministry of Education, Culture, Sports, Science and
Technology (MEXT) and a grant under the MEXT International
Science Research Program,\\
Japanese Society for the Promotion of Science (JSPS),\\
German Israeli Bi-national Science Foundation (GIF), \\
Bundesministerium f\"ur Bildung und Forschung, Germany, \\
National Research Council of Canada, \\
Hungarian Foundation for Scientific Research, OTKA T-038240, 
and T-042864,\\
The NWO/NATO Fund for Scientific Research, the Netherlands.\\

\end{document}